\UseRawInputEncoding
\documentclass[journal]{IEEEtran}
\usepackage{graphicx}
\usepackage{epstopdf}
\usepackage{amssymb}
\usepackage{float}
\usepackage{subfigure}
 \usepackage{cite}
 \usepackage{soul} % 导入 soul 包
\usepackage{color, xcolor} % 颜色包，color 必须导入，xcolor 建议导入
\usepackage{caption}
\usepackage{subfigure}
\usepackage{amsmath}
\usepackage{balance}
\usepackage{algorithm}
\usepackage{amsmath}
\usepackage{amsthm,amssymb}
\usepackage{mathrsfs}
\usepackage{algorithmic}
\usepackage{threeparttable}

\renewcommand{\algorithmicrequire}{\textbf{Input:}}
\renewcommand{\algorithmicensure}{\textbf{Output:}}

\ifCLASSINFOpdf
\begin{document}
% \title{Sampled-data practical tracking control for nonlinear time-delay systems }
\title{Differentially Private Distributed Nash Equilibrium Seeking over Time-Varying Digraphs}
%\sum

\author{ Ying~Chen, Qian~Ma%, \emph{Member, IEEE}
        %\emph{Fellow, IEEE}% <-this % stops a space
\thanks{This work was supported in part by NSFC under Grant 62173183 and 62473199. (Corresponding author: Qian Ma.)}
\thanks{Y. Chen and Q. Ma are with the  School of Automation, Nanjing University of Science and Technology, Nanjing 210094, China (e-mail: njcy2023@163.com; qma@njust.edu.cn).

}
% <-this % stops a space
}

\maketitle

\begin{abstract}
This paper proposes a new differentially private distributed Nash equilibrium seeking algorithm for aggregative games under time-varying unbalanced directed communication graphs. Random independent Laplace noises are injected into the transmitted information to protect players' sensitive information. Then, the push-sum consensus protocol is utilized to estimate the aggregate function with the perturbed information under the time-varying topologies. The weakening factor and the momentum term are designed to attenuate the negative affect of the noise and guarantee the convergence of the algorithm, respectively. The algorithm is then proven to ensure the almost sure convergence, as well as rigorous differential privacy with a finite cumulative privacy budget, without requiring a trade-off between provable convergence and differential privacy. Finally, the simulation is provided to demonstrate the effectiveness of the proposed algorithm.
\end{abstract}

\begin{IEEEkeywords}
Aggregative games, distributed Nash equilibrium seeking, differential privacy, time-varying digraphs.
\end{IEEEkeywords}

 \ifCLASSOPTIONpeerreview
 \begin{center} \bfseries EDICS Category: 3-BBND \end{center}
 \fi
\IEEEpeerreviewmaketitle

\section{Introduction}
\IEEEPARstart{A}{s} an important type of non-cooperative games, aggregative games, in which the payoff/cost function of each participating player depends on its own decision and the aggregate of all players' decisions, have attracted the attention of numerous scholars. Significant progress has been made in the research of distributed Nash equilibrium (NE) seeking algorithms for aggregative games \cite{JKo2015OR,FSa2016Auto,LPa2020TAC,DTA2022arViv,FangX2022TCS1,PF2020Auto,PF2020TCNS,CGu2024TAC}.

The distributed algorithms mentioned in the above works require direct sharing of estimated values or decision information in each iteration, which could potentially lead to the disclosure of sensitive individual information. Privacy preservation is particularly significant in non-cooperative games, as all players want to maximize their own benefits or minimize their own cost in the competition. In the last couple of years, efforts have been made to ensure the privacy preservation in aggregative games (see \cite{LuY2015IFAC,RCu2015ICWIE,ISh2021ORL} and the references therein). The design of these privacy preservation NE seeking algorithm is suitable for the case where a mediator exists. However, the presence of a mediator will increase the risk of privacy leakage. Therewith, the privacy preservation algorithm was designed for fully distributed NE seeking with the assistance of correlated noise in \cite{SGa2020IFAC}. And then a uncertain parameter approach was utilized to mask the pseudo-gradient mapping in \cite{MSh2022IJRNC}. However, both correlated noise and uncertain parameter restrict the applicability of the algorithm, and limit the privacy-preservation strength. There are some other methods to achieve privacy preservation in addition to adding noise approach, such as encryption and decomposition. Encryption is a widely embraced technique for protecting static datasets against unauthorized access. Cryptography-based approaches will also significantly increase communication and computation burden \cite{Ruan2019TAC}, which is not appropriate for systems with limited resources or systems with fast evolving behaviors or subject to hard real-time constraints. Decomposition is an alternative technique to preserve privacy without compromising data utility, however, it is difficult to evaluate and guarantee the strength of privacy protection \cite{Wang2019TAC1}.

Due to the strong robust protection against arbitrary post-processing and auxiliary information \cite{CDw2014FTTCS}, differential privacy has been widely applied in many issues to describe the privacy level, such as consensus control \cite{WangYM2023TC,Gaolan2021TC}, distributed optimization \cite{Xiongyy2020TCNS,MaoS2023TCS1}, and so on. For aggregative games problem, the recent works \cite{YEDP2021TAC} and \cite{WJM2022Auto} have proposed differential privacy fully distributed NE seeking algorithms. However, in order to guarantee rigorous $\epsilon$-differential privacy, these approaches have to compromise on the exact convergence to the NE. In order to circumvent the trade-off between convergence accuracy and rigorous $\epsilon$-differential privacy, the authors in \cite{WYQ2024TAC01} introduced a differential private distributed NE seeking algorithm for aggregative games on undirected graph. This algorithm was proved to achieve both rigorous differential privacy and converge almost surely to the NE with the assistance of a weakening factor sequence. The differential private distributed NE seeking under directed unbalanced communication graph was further developed in \cite{WYQ2024TAC02} to guarantee almost sure convergence as well as rigorous differential privacy with summable cumulative privacy budget. Furthermore, in the problem of privacy preservation for games, communication topology structure is a key factor influencing the distributed NE seeking. The authors in \cite{YEDP2021TAC} designed the differential private NE seeking algorithm considering time-varying undirected topology with the help of the symmetry of communication graph assuming a doubly stochastic weight matrix. However, if the time-varying graphs are directed and unbalanced, then assuming a weight matrix that is doubly stochastic will be challenging. Meanwhile, the symmetry of the graph cannot be guaranteed, which will bring great difficulty in the effectiveness analysis of the algorithm. The literature \cite{WYQ2024TAC02} developed the distributed differential privacy NE seeking algorithm based on the left eigenvector of weight matrix for fixed directed unbalanced graph. The left eigenvector is associated with the weight matrix of the fixed graph, which is determined by the weight matrix of the global communication graph. If the communication topology experiences time-varying changes, then the method in \cite{WYQ2024TAC02} will be no longer applicable. Therefore, designing new NE seeking algorithms for more general time-varying unbalanced directed graphs in aggregative games, while achieving exact convergence to the NE and ensuring rigorous differential privacy, remains a significant challenge.

Inspired by the discussion above, this paper explores the problem of distributed NE seeking with differential privacy for aggregative games on time-varying unbalanced directed graphs. The random Laplace noise is injected into the information transmitted among players to avoid the real information being leaked. The push-sum consensus protocol is used to estimate the aggregate function for overcoming the difficulty of being unable to utilize the symmetry and balance of the weight matrix caused by time-varying unbalanced directed graphs. To mitigate the adverse impact of the differential privacy noise on the convergence of the algorithm, the weakening factor and the momentum term are designed. The proposed algorithm ensures almost sure convergence, as well as rigorous differential privacy with a summable cumulative privacy budget, without the trade-off between the exact convergence and rigorous differential privacy.

The key contributions of this paper are as follows. First, a more general communication topology model is considered for differential private distributed NE seeking problem compared to the models in references \cite{YEDP2021TAC}, \cite{WYQ2024TAC02}. Second, the more general push-sum method is employed in our paper to handle the complex topologies, which means that our proposed approach can be applied not only to time-varying unbalanced directed topologies considered in this study, but also more easily to the common fixed undirected graph and directed balanced graph satisfying certain conditions. Third, the stringent requirement of doubly stochastic weight matrix as stated in references \cite{YEDP2021TAC} and \cite{WJM2022Auto} is removed. Instead, a more relaxed column stochastic matrix is used in this paper. The last not not least, the aggregate function in \cite{YEDP2021TAC} and \cite{WYQ2024TAC02}, which is a direct summation of strategies, is extended to the more general function, which increases the flexibility of aggregate function design.

\section{Preliminaries and Problem description}
\subsection{ Graph Theory}
 A time-varying directed graph $\mathcal{T}(l)=(\mathcal{X}, \mathcal{S}(l), B(l))$ with the node set $\mathcal{X}=\{1, {\dots }, N\}$, and the edge set $\mathcal{S}(l)$ at the time $l$, is considered. $(j, i) \in \mathcal{S}(l)$ denotes that there is an edge from node $j$ to node $i$, which means that the $i$-th player can receive the information of $j$-th player directly at time $l$. Define in-neighbor sets of $i$-th as $N_i^{-}(l)$, which is $\{j|(j,i) \in \mathcal{S}(l)\} \bigcup \{ i \} \}$. Define out-neighbor sets of $i$-th as $N_i^{+}(l)$, which is $\{j|(i,j) \in \mathcal{S}(l)\} \bigcup \{ i \} \}$.  A fixed digraph $\mathcal{T}$ is strongly connected if there exists a directed path from any player $i$ to any other distinct player $j$. For the time-varying digraph $\mathcal{T}(l)$, the sequence $\{\mathcal{T}(l)\}$ is said to be $D$-strongly connected if there exists a $D \in \mathbb{Z}_{+}$ such that the union digraph $\bigcup_{r=lD}^{(l+1)D-1}\mathcal{T}(r)$ is strongly connected for any $l \in \mathbb{Z}_{+}$. $B(l)$ is weight matrix whose $(i,j)$-th entry is $B_{ij}(l)$ at the $l$-th iteration, with $B_{ij}(l)>0$ if $j \in N_i^{-}(l)$, $B_{ij}(l)=0$ otherwise. Note that $\mathbb{R}$ represents the set of real numbers, and $\mathbb{Z}_{+}$ represents the set of positive integers.

\subsection{Problem Description of Games}
An aggregative problem involving $N$ players is considered. Each player $i$ strives to minimize its private cost function $J_{i}(q): U{\rightarrow} \mathbb{R}$ by altering its personal action $q_i \in U_i$, which is described as follows:
\begin{equation}
\mathop{\text{min}}\limits_{q_i}J_{i}(q_i,{\sigma}(q)),\ {\sigma}(q)=\frac{1}{N}\sum_{i = 1}^N \phi_i(q_i), \quad \forall i \in \mathcal{X},   \label{obiectFunc1}
\end{equation}
where $q=[q_1,q_2, \ldots, q_N]^T$. $U=U_1{\times}\cdots {\times} U_N$, $U_i$ is the action set for $i$-th player. ${\sigma}(q)$ denotes the aggregate of all players' actions. $\phi_i(q_i)$ is a local map only known to player $i$. Note that $\nabla J(q)=\text{col}\{{\nabla}_{q_i}J_{i}(q_1,{\sigma}(q)),\cdots,{\nabla}_{q_i}J_{i}(q_N,{\sigma}(q))\}$, where col$\{\cdot\}$ is a column vector. The partial derivative of function $J_i(q_i(l),q_{-i}(l))$ with respect to $q_i$ is $g_{i}(q_{i}(l),{\sigma}(q))$$=$${\nabla}_{q_i}J_i(q_i(l),q_{-i}(l))$$=$${\nabla}_{q_i}J_{i}(q_i(l),{\sigma}(q)) +\frac{1}{N}{\nabla}_{\sigma(q)}J_{i}(q_i(l),{\sigma}(q))\nabla_{q_i}\phi_i(q_i)$, where $q_{-i}(l)=[q_{1}(l),\cdots,q_{i-1}(l),q_{i+1}(l)\cdots,q_{N}(l)]^T$. Then the partial derivative of function $J_i(q_i(l),y_i(l))$ with respect to $q_i$ can be obtained:
\setlength\arraycolsep{1pt}
\begin{eqnarray}
g_{i}(q_{i}(l),y_i(l))&=&{\nabla}_{q_i}J_{i}(q_i(l),{\sigma}(q))|_{{\sigma}(q)=y_i(l)} \nonumber \\
&&+\frac{1}{N}{\nabla}_{\sigma(q)}J_{i}(q_i(l),{\sigma}(q))\nabla_{q_i}\phi_i(q_i)|_{{\sigma}(q)=y_i(l)} ,\nonumber
\end{eqnarray}
where $l$ denotes $l$-th iteration.

\emph{Remark 1.} Aggregative games are powerful theoretical models for describing competitive scenarios characterized by aggregative interactions among multiple players. In many practical situations, each player's objective function depends not only on their own strategy but also on the aggregate actions of all players. For instance, in the energy consumption model outlined in \cite{YE2017TC}, the electricity consumption of users is influenced by their individual energy usage as well as the total energy consumption within the market. Similarly, in a manufacturing enterprise, the output of each factory depends on both its own production and the average output of all engaged factories \cite{Wang2014ICT}. In model (\ref{obiectFunc1}), as a linear function for the local contribution to the aggregate function, $\phi_i(q_i)$ can capture the characteristics and needs of different individuals. This means that each player's decision and contribution can be adjusted based on the specific circumstances, thereby can reflect the complex behaviors in reality more accurately.

NE is a solution concept for the aggregative games problem (\ref{obiectFunc1}) where each player's action reaches a strategy profile, and no player can reduce their own cost by unilaterally changing their decision. The definition of NE is as stated below:

\emph{Definition 1 \cite{YEDP2021TAC}.} Considering a game $\overline{H}=\{\mathcal{X},U,\mathbb{J}\}$, the strategy profile $q^{*}=(q_{1}^{*},\ldots ,q_{N}^{*})^{T} \in U$  is an NE of the game $\overline{H}$ if $J_{i}\bigl(q_{i}^{*},q_{-i}^{*} \bigr)\leq J_{i}\bigl(q_{i},q_{-i}^{*} \bigr), \forall i \in \mathcal{X}$, where $q_i \in U_i$, $q_{-i}=(q_{1}, \ldots, q_{i-1}, q_{i+1}, \ldots, q_{N})^{T} \in U_1 \times \cdots \times U_{i-1} \times U_{i+1} \times \cdots \times U_N$, $\mathbb{J}\triangleq \{ J_1, \cdots J_N\}$.

\emph{Assumption 1.} There exists $D>0$ to satisfy that the time-varying directed graph $\mathcal{T}(l)$ is $D$-strongly connected.

\emph{Assumption 2.} For all $i \in \mathcal{X}$, $U_i$ is nonempty, compact and convex. $J_{i}(q)$ and $\phi_i(q_i)$ are continuously differentiable on $q$ and $q_i$, respectively. Moreover, for given $q_{-i} \in U_1 \times \cdots \times U_{i-1} \times U_{i+1} \times \cdots \times U_N$, $J_{i}(q_i, q_{-i})$ is convex about $q_i$ on $U_i$.

\emph{Assumption 3.} The mapping $\nabla J(q): U{\rightarrow}\mathbb{R}$ is strongly monotone on $U$, i.e., there exists a constant $m>0$ such that for $\forall q,q' \in U, q{\ne}q'$, there has
$(\nabla J(q)-\nabla J(q'))^T(q-q')>m\Vert q-q' \Vert^2_2$, where $\Vert \cdot \Vert_2$ is Euclidean norm.

\emph{Assumption 4.} The mapping $g_{i}(q_{i}(l),{\sigma})$ is $L_{i_1}$-Lipschitz continuous for ${\sigma}(q) \in \mathbb{R}$, and ${\phi}_i(q_i)$ is $L_{i_2}$-Lipschitz continuous for $q_i \in U_i$, i.e.,
\begin{eqnarray}
\vert\vert g_{i}(q_{i}(l),{\sigma}_1)-g_{i}(q_{i}(l),{\sigma}_2)\vert\vert_2 &\le& L_{i_
1} \vert\vert \sigma_1-\sigma_2 \vert\vert_2, \ \forall \sigma_1,\sigma_2 \in \mathbb{R} \nonumber \\
\vert\vert {\phi}_i(q_1)-{\phi}_i(q_2)\vert\vert_2 &\le& L_{i_2} \vert\vert q_1-q_2 \vert\vert_2, \ \forall q_1,q_2 \in U_i \nonumber
\end{eqnarray}
where $L_{i_1}>0$, $L_{i_2}>0$.

\emph{Lemma 1 \cite{AN2015TAC}}. Let ${\mu(l)}$ be a scalar sequence. The following results can be obtained: \\
\noindent(1) If $\mathop{\lim}_{l\to\infty}\mu(l)=\mu$, and $0<\beta<1$, then $\mathop{\lim}_{l\to\infty}\sum_{r = 0}^l \beta^{l-r}\mu(r)=\frac{\mu}{1-\beta}$, for $\forall \ l \ge 0$ and $\forall \ \mu(0)< \infty$.

\noindent(2) If $\mu(l) \ge 0 $ for all $l$, $\sum_{l = 0}^{\infty} \mu(l)<\infty$, and $0<\beta<1$, then $\sum_{l = 0}^{\infty}(\sum_{r = 0}^l\beta^{l-r}\mu(r))<\infty$.

\emph{Lemma 2 \cite{MaoS2023TCS1}.} If directed graph $\mathcal{T}(l)$ satisfies Assumption 1, then there is a stochastic vector ${\Psi}(l)$, $0 \le r \le l$ such that
\begin{eqnarray}
\Vert [B(l)B(l-1)\cdots B(r+1)B(r)]_{ij}- {\Psi}_i(l)\Vert \le C_1{\lambda}^{l-r}, \nonumber
\end{eqnarray}
where $C_1>0$, $\lambda \in (0,1)$. $B(l)=[B_{ij}(l)]$ with $B_{ij}(l)$ being the $i$-th row and $j$-th column element of the matrix $B(l)$.

\emph{Remark 2.} From \cite{AN2015TAC}, $[B(l:0)\mathbf{1}_N]_i>\bar{\delta}$ can be obtained, where $\bar{\delta}$ is the positive constant, $\mathbf{1}_N$ is $N$-dimensional vector. This means that the row sum of $B(l:0)$ is at least $\bar{\delta}$, and the parameter $\bar{\delta}$ can be seen as a measure of imbalance of the influence among the nodes.

\subsection{ Differential Privacy}
Injecting Laplace noise into the transmitted information is a popular method for achieving differential privacy. $x \sim Lap(b(l))$ is used to denote the Laplace distribution of a random variable $x$ with the probability density function $\frac{1}{2b}e^{-\frac{|x|}{b(l)}}$ at the time $l$, where $b(l)>0$. It can be obtained that $Lap(b(l))$ has zero mean and variance $2b^2(l)$. Then adjacency of two distributed NE seeking problems is defined as follows.

\emph{Definition 2 \cite{WYQ2024TAC01}.} Two distributed NE seeking problems $\widetilde{H}=\{U,\mathbb{J},B(l)\}$ and $\widetilde{H}'=\{U',\mathbb{J'},B'(l)\}$ are considered adjacent if the following three conditions are met:

\noindent(1) The domains of decision variables $U=U'$ are identical, and the domains of the interaction weight matrices $B(l)=B'(l)$ are identical at the $l$-th iteration, $\forall l>0$;

\noindent(2) $\mathbb{J}$ and $\mathbb{J'}$ are two different cost function sets with $\mathbb{J}\triangleq \{ J_1, \cdots J_N\}$. Only one player's cost function is different, i.e., there exists an $i \in \mathcal{X}$ such that $J_i \ne J_i^{'}$, and $J_j = J_j^{'}, \forall j \in \mathcal{X}$ and $j \ne i$;

\noindent(3) The distinct cost functions $J_i$ and $J_i^{'}$ exhibit similar characteristics near $q^{*}$, the NE of $\widetilde{H}$. To be specific, there exits some $o>0$ such that for all $q=$col$(q_1,\cdots,q_N)$ and $q'=$col$(q'_1,\cdots,q'_N)$ in $\tilde{B}_{\iota}\triangleq\{u:u \in U, \Vert u-q^{*}\Vert < o\}$, we have $\nabla J_{i}(q_i,{\sigma})=\nabla J'_{i}(q'_i,{\sigma}')$, where ${\sigma}=\frac{1}{N}\sum_{i = 1}^N \phi_i(q_i)$, ${\sigma}'=\frac{1}{N}\sum_{i = 1}^N \phi_i(q'_i)$.

Here we abbreviate ${\sigma}(q)$ to $\sigma$ and abbreviate ${\sigma}(q')$ to $\sigma'$.

For a distributed NE seeking algorithm, $\mathcal{A}$ is used to denote an infinite execution sequence of the designed algorithm, which is an iteration variable sequence $\mathcal{A}=\{\tau(0), \tau(1),\cdots\}$, and $\tau(l)=\{q_i(l),\hat{\sigma}_i(l),\hat{w}_i(l),z_i(l)\}, \ l=0,1,\cdots, \ i=1,\cdots,N$, is state set generated by algorithm at $l$-th iteration with $q_i(l),\hat{\sigma}_i(l),\hat{w}_i(l),z_i(l)$ defined in algorithms (\ref{algorithm01si})-(\ref{algorithm01xi}) in part III of our paper. Adversaries are assumed to be able to observe all information $\mathcal{O}$ transmitted in the player's communication channel, and $\mathcal{O}=\{s_i(0),s_i(1),\cdots\}, \ l=0,1,\cdots, \ i=1,\cdots,N$. For a given distributed NE seeking problem $\widetilde{H}$ with an initial state $\tau(0)$, the observation mapping function is defined as $\mathcal{R}_{\widetilde{H},\tau(0)}(\mathcal{A})\triangleq \mathcal{O}$, i.e., $\mathcal{R}_{\widetilde{H},\tau(0)}: \mathcal{A} \rightarrow \mathcal{O}$. And the set of executions $\mathcal{A}$ of $\mathcal{O}$ is defined as $\mathcal{R}^{-1}_{\widetilde{H},\tau(0)}(\mathcal{O})$, i.e., $\mathcal{R}_{\widetilde{H},\tau(0)}^{-1}:\mathcal{O} \rightarrow  \mathcal{A}$.

Then the concept of $\epsilon$-differential privacy is given as follows.

\emph{Definition 3 \cite{WYQ2024TAC01} ($\epsilon$-differential privacy).} For a given $\epsilon > 0$, if for any two adjacent distributed NE seeking problems $\widetilde{H}$ and $\widetilde{H}'$, any set of observation sequences $\mathcal{O}_s \subseteq \mathbb{O}$, and any initial state $\tau(0)$, the following relationship always holds:\ \ $\mathbb{P}[\mathcal{R}_{\widetilde{H},\tau(0)} \in \mathcal{O}_s] \le e^{\epsilon}\mathbb{P}[\mathcal{R}_{\widetilde{H}',\tau(0)} \in \mathcal{O}_s]$, then an iterative distributed algorithm is $\epsilon$-differential privacy. Here, $\mathbb{P}$ denotes the probability. $\mathbb{O}$ is the set of all possible observation sequences defined by $\mathbb{O}=\mathop{\text{U}}\limits_{t \in T}\{ \mathcal{O}_t \vert  \mathcal{O}_t \in \mathcal{F}_t \}$, where $\mathcal{O}_t$ is the information observed at time $t$, $T$ is the set of all $t$, $\mathcal{F}_t$ is the $\sigma$-algebra generated by the history of an algorithm up to time $t$.

$\epsilon$-differential privacy in Definition 3 ensures that an adversary having access to all shared messages cannot gain information with a significant probability of any participating player's cost function. It can be seen that a smaller $\epsilon$ means a higher level of privacy preservation. The players' objective functions are considered to be sensitive information to be protected in our paper.

Our objective is to develop a differential private distributed NE seeking algorithm for the aggregative games problem (\ref{obiectFunc1}) under time-varying unbalanced digraphs such that the players' actions will achieve almost sure convergence to the NE, and the algorithm can achieve the rigorous $\epsilon$-differential privacy with finite cumulative privacy budget.

\section{Differential Private Distributed NE seeking algorithm design}

Our differential private distributed NE seeking algorithm is developed in Algorithm 1.
\begin{algorithm}
	\renewcommand{\thealgorithm}{}
    \floatname{algorithm}{Algorithm 1.}
    \newcommand{\blue}[1]{#1}
	\renewcommand{\algorithmicrequire}{\quad \textbf{Initialization:}}
    \renewcommand{\algorithmicensure}{\quad \textbf{Parameters:}}\
	\caption{Differential Private Distributed NE seeking algorithm for aggregative games}
\blue{\begin{algorithmic}
    \ENSURE Stepsize $\mu_{i}(l)>0$, weakening factor $\rho(l)>0$, momentum coefficient $\beta_{i}>0$.
	\REQUIRE $\hat{w}_i(0)=1$, $q_{i}(0)=q_{i}(1)$, $\hat{\sigma}_i(0)=q_{i}(0)$.
%\resetindent % 重置缩进
  \FOR{$l=1,2,\dots$}
        \STATE \textbf{Step 1:} Every player $j$ adds the differential-privacy noise $\varepsilon_j(l)$ to its estimate $\hat{\sigma}_j(l)$, and then sends the disturbed estimate $s_j(l)$ to player $i \in N_j^{+}(l)$.
        \begin{eqnarray}
        s_j(l) &=& \hat{\sigma}_j(l) + \varepsilon_j(l), \label{algorithm01si}
        \end{eqnarray}
        \STATE \textbf{Step 2:} After receiving $s_j(l)$ from all $j \in N_i^{-}(l)$, the player $i$ execute the following update algorithms:
\begin{eqnarray} % 原始算法
\hat{w}_i(l+1)&=&\sum_{i = 1}^N B_{ij}(l)\hat{w}_j(l), \label{algorithm01wi} \\
\hat{\sigma}_i(l+1) &=& \rho(l)z_i(l+1)+ \phi_i(q_i(l+1))-\phi_i(q_i(l)), \ \ \label{algorithm01sigma} \\
z_i(l+1)&=&\sum_{i = 1}^N B_{ij}(l)s_j(l), \label{algorithm01zi} \\
y_i(l+1)&=&\frac{\rho(l)z_i(l+1)}{\hat{w}_i(l+1)}, \label{algorithm01yi}\\
q_{i}(l+1)&=&P_{U_i}[ q_{i}(l)-\mu_{i}(l)g_{i}(q_{i}(l),y_i(l)) \nonumber \\
&&+\beta_{i}(q_{i}(l)-q_{i}(l-1)) ],   \label{algorithm01xi}
\end{eqnarray}
\ENDFOR
\end{algorithmic}}
\end{algorithm}

In Algorithm 1, $\hat{w}_i(l)$ and $y_i(l)$ are auxiliary variables to balance the influence of unbalanced directed graph. $\beta_{i}(q_{i}(l)-q_{i}(l-1))$ is the momentum term. $B_{ij}(l)=\frac{1}{|N_j^{+}(l)| }$ if $j \in N_i^{-}(l)$, otherwise, $B_{ij}(l)=0$. $|N_j^{+}(l)|$ denotes the number of out-neighbor of the $j$-th player. $\sum_{i = 1}^NB_{ij}(l)=1$. The noise $\varepsilon_i(l)\sim Lap(b_i(l))$, $\mathbb{E}[\varepsilon_i(l)]=0$, $\mathbb{V}[\varepsilon_i(l)]=2(b_i(l))^2$, where $\mathbb{V}[\cdot]$ denotes the symbol of variance. Then $\mathbb{E}[|\varepsilon_i(l)|] \le \sqrt{2}b_i(l)$ can easily be obtained. $P_{s}[\cdot]$ is the projection of a vector onto the set $s$. And $\mathcal{F}_l$ in our paper is defined as $\mathcal{F}_l=\{ q_i(k),\hat{\sigma}_i(k),\hat{w}_i(k),z_i(k),\ i=1,\cdots,N, \ 0 \le k\le l \}$.

\emph{Remark 3.} When designing differential privacy algorithms, the choice of noise distribution, such as Gaussian distribution or Laplace distribution, is indeed of importance. As a matter of fact, Gaussian noise can also be employed to achieve a given differential privacy budget $\epsilon_l$ at the iteration $l$ by replacing the $\ell_1$ norm ($\Vert \cdot \Vert_1$) in the sensitivity Definition 4 of section V with the $\ell_2$ norm ($\Vert \cdot \Vert_2$) (see Appendix A of \cite{Dwork2014Book}). However, although Gaussian distribution appears to be more suitable due to its excellent performance in many linear dynamic systems, Laplace noise is the earliest noise model proposed for achieving differential privacy \cite{Dwork2006Book}, and it can achieve a more strict form of differential privacy than Gaussian distribution. Furthermore, the mathematical properties of Laplace noise simplify the analysis of algorithmic complexity and the proofs of privacy guarantees.

\emph{Remark 4.} For the above algorithm, $q_{i}(l)$ is the $i$-th player's action. $\hat{\sigma}_i(l)$ is the estimation of $i$-th player for the aggregate function $\sigma(q)$. And the local function $\phi_i(q_i(l))$ used in the aggregate function $\sigma(q)$ is more universal than that $q_i(l)$ in \cite{YEDP2021TAC}, \cite{WYQ2024TAC01,WYQ2024TAC02}. The momentum term $\beta_{i}(q_{i}(l)-q_{i}(l-1))$ in (7) is motivated by the heavy-ball method \cite{Polyak1964}. By introducing the momentum term into the gradient descent method, the heavy-ball method uses the current iteration and the previous one to calculate the next so as to accelerate the convergence, which can be seen that the convergence is faster than that of the literature \cite{WYQ2024TAC01}. More details on the distributed heavy-ball methods can be found in \cite{XinR2020TAC}. $\{\rho(l)\}$, as the weakening factor sequence, is used for weakening the effect of noise. To achieve strong differential privacy, independent noise should be injected repeatedly in every round of message sharing and, hence, constantly affecting the algorithm through inter-player interactions and leading to significant reduction in algorithmic accuracy. Thus, the weakening factor sequence is utilized to gradually weaken inter-player interactions, which guarantees that the influence of differential-privacy noise on computation accuracy is reduced, and the convergence to the NE is ensured even in the presence of persistent differential-privacy noise. Maybe the better ways without weakening factor or the stronger assumption to deal with the negative affect of noise can be explored in future work.

Define $z(l)=[z_1(l),\cdots,z_N(l)]^T$, $\hat{\sigma}(l)=[\hat{\sigma}_1(l),\cdots,\hat{\sigma}_N(l)]^T$, $\hat{w}(l)=[\hat{w}_1(l),\cdots,\hat{w}_N(l)]^T$, $\varepsilon(l)=[\varepsilon_1(l),\cdots,\varepsilon_N(l)]^T$, $\phi(l)=[\phi_1(l),\cdots,\phi_N(l)]^T$, then from (\ref{algorithm01wi}), (\ref{algorithm01sigma}), (\ref{algorithm01zi}) and (\ref{algorithm01yi}), the following compact form can be obtained:
\begin{eqnarray} % 原始算法
y_i(l+1)&=&\frac{\rho(l)z_i(l+1)}{\hat{w}_i(l+1)}, \label{algorithm01yi02} \\
\hat{w}(l+1)&=&B(l)\hat{w}(l), \label{algorithm01wi02} \\
z(l+1)&=&B(l)(\hat{\sigma}(l) + \varepsilon(l)), \label{algorithm01zi02} \\
\hat{\sigma}(l+1) &=& \rho(l)z(l+1)+ \phi(l+1)-\phi(l), \label{algorithm01sigma02}
\end{eqnarray}
here we abbreviate $\phi_i(q_i(l))$ to $\phi_i(l)$ and abbreviate $\hat{\sigma}_i(q_i(l))$ to $\hat{\sigma}_i(l)$.

Some other assumptions need to be given for facilitating the following proof in next part.

\emph{Assumption 5.} Step-size $\mu_i(l)$ satisfies the following conditions:

\noindent(1) $0<\mu_i(l)<1$ for $\forall l \in \mathbb{N}_+, i \in \mathcal{X}$, and each $\mu_i(l)$ is monotonically non-increasing;

\noindent(2) $\sum_{l = 1}^{\infty}\mu_i(l) < \infty$, $\sum_{l = 1}^{\infty}( \mu_i(l) )^2 < \infty$;

\noindent(3) $\sum_{l = 0}^{\infty}( \bar{\mu}(l)- \underline{\mu}(l)) <\infty$, where $\bar{\mu}(l)=\mathop{\text{max}}\limits_{j \in \mathcal{X}} \{ \mu_j(l) \}$, and $\underline{\mu}(l)=\mathop{\text{min}}\limits_{j \in \mathcal{X}} \{ \mu_j(l) \}$;

\noindent(4) $\sum_{l = 1}^{\infty} (\frac{\mu_i(l)}{\rho(l)})^2 < \infty$.

\emph{Assumption 6.} The weakening factor $\rho(l)$ satisfies the following conditions:

\noindent(1) $0<\rho(l+1) \le \rho(l)<1$;

\noindent(2) $\sum_{l = 0}^{\infty}\rho(l) < \infty$;

\noindent(3) $\sum_{l = 0}^{\infty}(\rho(l)\check{b}(l)) < \infty$, where $\check{b}(l)=\mathop{\text{max}}\limits_{i \in \mathcal{X}} \{b_i(l)\}$.

\emph{Remark 5.} Assumption 5 and Assumption 6 can be satisfied simultaneously to some extent if $\mu_i(l) \in (0,1)$, $\rho(l)\in (0,1)$, and the rate of increase of $\check{b}(l)$ is slower than the rate of decrease of $\rho(l)$. For example, $\mu_{i}(l)=\frac{1}{1+0.0001* 2^{0.01l+2}}$, $\rho(l)=\frac{1}{1+0.1*l^{2.01}}$, $\check{b}(l)=l+2$, then condition (2) of Assumption 5 and condition (1) of Assumption 6 are satisfied. Calculating by summing a sequences $\sum_{l = 1}^{\infty} \frac{1}{1+0.0001* 2^{0.01l+2}}$, $\sum_{l = 1}^{\infty} (\frac{1}{1+0.0001* 2^{0.01l+2}})^2$, $\sum_{l = 1}^{\infty} (\frac{1+0.1*l^{2.01}}{1+0.0001* 2^{0.01l+2}})^2$, $\sum_{l = 1}^{\infty}\frac{1}{1+0.1*l^{2.01}}$, and $\sum_{l = 1}^{\infty} \frac{l+2}{1+0.1*l^{2.01}}$, we can obtain that all conditions of Assumption 5 and Assumption 6 are satisfied, and the five summation sequences approach $1.42857 \times 10^6$, $7.52\times10^9$, $4.948774565 \times 10^{12}$, $16.27$ and $991.5$, respectively.

\section{CONVERGENCE ANALYSIS OF ALGORITHM}
This section will prove the convergence of the proposed algorithm for aggregative games on time-varying digraphs.

For convenience, we define the following notations:

\noindent (1) $B(l:r) \triangleq B(l)B(l-1) \cdots B(r), \ l > r \ge 0$; if $l<r$, $B(l:r) \triangleq I_N$; if $l=r$, $B(l:r) \triangleq B(l)$.

\noindent(2) $H(l,s)=B(l:r)-{\Psi}(l) \mathbf{1}_N^T,  \ l > r \ge 0$; if $l<r$, $H(l,r) \triangleq I_N$; if $l=r$, $H(l,r) \triangleq H(l)$.

\noindent(3) $\mathop{\Pi}\limits_{m=r}^l\rho(m)=\rho(l)\rho(l-1)\cdots\rho(r), \ l \ge r \ge 0$; if $l<r$, $\mathop{\Pi}\limits_{m=r}^l\rho(m)=1$.

Some lemmas need to be given for getting the final convergence result of the designed algorithms.

\emph{Lemma 3.} If Assumptions 1-6 hold, then the following inequality can be derived:
\begin{eqnarray}
\mathbb{E}\Big[ \Vert y_i(l+1)-\frac{\mathbf{1}_N^T \hat{\sigma}(l+1)}{N} \Vert \big\vert \mathcal{F}_l\Big] \le C_5, \label{Eysigma01}
\end{eqnarray}
where $C_5>0$. $\mathcal{F}_l$ will be omitted in the following text.

\emph{Proof: } Define ${\Delta}\phi(l)=\phi(l)-\phi(l-1)$, ${\Delta}\phi_i(l)=\phi_i(l)-\phi_i(l-1)$. According to (\ref{algorithm01yi02})-(\ref{algorithm01sigma02}), the equality below can be obtained
\setlength\arraycolsep{2pt}
\begin{eqnarray}
\hat{\sigma}(l+1) &=& \big(\mathop{\Pi}\limits_{m=0}^l\rho(m)\big) B(l:0)\hat{\sigma}(0) \nonumber \\
&&+\sum_{r = 0}^{l} \big(\mathop{\Pi}\limits_{m=r}^l\rho(m)\big) B(l:r)\varepsilon(r) \nonumber \\
&&+\sum_{r = 1}^{l+1} \big(\mathop{\Pi}\limits_{m=r}^l\rho(m)\big)B(l:r){\Delta}\phi(r). \  \quad \label{algorithm01sigma03}
\end{eqnarray}

Then, multiplying both sides of formula (\ref{algorithm01sigma03}) by $\mathbf{1}_N^T$ and $B(l+1)$, respectively, one can get
\setlength\arraycolsep{2pt}
\begin{eqnarray}
&&\mathbf{1}_N^T\hat{\sigma}(l+1)\nonumber \\
&=& \big(\mathop{\Pi}\limits_{m=0}^l\rho(m)\big) \mathbf{1}_N^T\hat{\sigma}(0)  + \sum_{r = 0}^{l} \big(\mathop{\Pi}\limits_{m=r}^t\rho(m)\big)\mathbf{1}_N^T\varepsilon(r)  \nonumber \\
&&+\sum_{r = 1}^{l+1} \big(\mathop{\Pi}\limits_{m=r}^l\rho(m)\big)\mathbf{1}_N^T{\Delta}\phi(r).  \label{algorithm01sigma04}
\end{eqnarray}

Multiplying both sides of formula (\ref{algorithm01sigma03}) by $B(l+1)$
\setlength\arraycolsep{2pt}
\begin{eqnarray}
B(l+1)\hat{\sigma}(l+1) &=& \big(\mathop{\Pi}\limits_{m=0}^l\rho(m)\big) B(l+1:0)\hat{\sigma}(0) + \nonumber \\
&& \sum_{r = 0}^{l} \big(\mathop{\Pi}\limits_{m=r}^l\rho(m)\big)B(l+1:r)\varepsilon(r)  + \nonumber \\
&&\sum_{r = 1}^{l+1} \big(\mathop{\Pi}\limits_{m=r}^l\rho(m)\big)B(l+1:r){\Delta}\phi(r). \quad  \quad \label{algorithm01sigma0501}
\end{eqnarray}

Henceforth, from (\ref{algorithm01sigma04}) and (\ref{algorithm01sigma0501}), we can know
\begin{eqnarray}
&&B(l+1)\hat{\sigma}(l+1)-{\Psi}(l+1)\mathbf{1}_N^T\hat{\sigma}(l+1) \nonumber \\
&=& \big(\mathop{\Pi}\limits_{m=0}^l\rho(m)\big) H(l+1,0)\hat{\sigma}(0)  \nonumber \\
&&+\sum_{r = 0}^{l} \big(\mathop{\Pi}\limits_{m=r}^l\rho(m)\big)H(l+1,r)\varepsilon(r) \nonumber \\
&&+\sum_{r = 1}^{l+1} \big(\mathop{\Pi}\limits_{m=r}^l\rho(m)\big)H(l+1,r){\Delta}\phi(r).  \label{algorithm01sigma06}
\end{eqnarray}

Subsequently, based on (\ref{algorithm01zi02}) and (\ref{algorithm01sigma06}), the following result can be obtained
\begin{eqnarray}
z(l+1)&=&B(t)(\hat{\sigma}(l) + \varepsilon(l)) \nonumber \\
&=& {\Psi}(l)\mathbf{1}_N^T\hat{\sigma}(t)+B(l)\varepsilon(l)\nonumber \\
&&+\big(\mathop{\Pi}\limits_{m=0}^{l-1}\rho(m)\big) H(l,0)\hat{\sigma}(0) \nonumber \\
&&+\sum_{r = 0}^{l-1} \big(\mathop{\Pi}\limits_{m=r}^{l-1}\rho(m)\big)H(l,r)\varepsilon(r) \nonumber \\
&&+\sum_{r = 1}^{l} \big(\mathop{\Pi}\limits_{m=r}^{l-1}\rho(m)\big)H(l,r){\Delta}\phi(r).
\label{algorithm01zi03}
\end{eqnarray}

Then multiplying both sides of formula (\ref{algorithm01zi03}) by $\rho(l)$, we can acquire
\begin{eqnarray}
\rho(l)z(l+1)&=&\rho(l){\Psi}(t)\mathbf{1}_N^T\hat{\sigma}(l)+\rho(l)B(l)\varepsilon(l)\nonumber \\
&&+\big(\mathop{\Pi}\limits_{m=0}^{l}\rho(m)\big) H(l,0)\hat{\sigma}(0) \nonumber \\
&&+\sum_{r = 0}^{l-1} \big(\mathop{\Pi}\limits_{m=r}^{l}\rho(m)\big)H(l,r)\varepsilon(r) \nonumber \\
&&+\sum_{r = 1}^{l} \big(\mathop{\Pi}\limits_{m=r}^{l}\rho(m)\big)H(l,r){\Delta}\phi(r). \label{algorithm01yi0301}
\end{eqnarray}

In light of (\ref{algorithm01wi02}) and $\hat{w}_i(0)=1$, $\hat{w}(l+1)$ can be derived
\begin{eqnarray}
\hat{w}(l+1)&=&B(l)\hat{w}(l) =H(l:0)\mathbf{1}_N+{\Psi}(l)N. \label{algorithm01wi03}
\end{eqnarray}

Then according to (\ref{algorithm01yi02}), (\ref{algorithm01yi0301}) and (\ref{algorithm01wi03}), we can get
\begin{eqnarray}
y_i(l+1)&=&\Big([\rho(l){\Psi}(l)\mathbf{1}_N^T\hat{\sigma}(l)+\rho(l)B(l)\varepsilon(l)\nonumber \\
&&+\big(\mathop{\Pi}\limits_{m=0}^{l}\rho(m)\big) H(l,0)\hat{\sigma}(0) \nonumber \\
&&+\sum_{r = 0}^{l-1} \big(\mathop{\Pi}\limits_{m=r}^{l}\rho(m)\big)H(l,r)\varepsilon(r) \nonumber \\
&&+\sum_{r = 1}^{l} \big(\mathop{\Pi}\limits_{m=r}^{l}\rho(m)\big)H(l,r){\Delta}\phi(r)]_i \Big)\Big/ \nonumber \\
&&\Big([H(l:0)\mathbf{1}_N]_i+{\Psi}_i(l)N\Big).
\label{algorithm01yi03}
\end{eqnarray}

According to (\ref{algorithm01zi02}) and (\ref{algorithm01sigma02}), the equation provided below stands
\setlength\arraycolsep{1pt}
\begin{eqnarray}
\frac{\mathbf{1}_N^T}{N}\hat{\sigma}(l+1) &=& \frac{\mathbf{1}_N^T}{N}\rho(l)\sigma(l)+ \frac{\mathbf{1}_N^T}{N}\rho(l)+ \frac{\mathbf{1}_N^T}{N} \Delta \phi(l+1), \ \quad  \quad \label{algorithm01sigma07}
\end{eqnarray}
where $\sigma(l)=[\sigma_1(l),\cdots,\sigma_N(l)]^T$. Here we abbreviate $\sigma_i(q_i(l))$ to $\sigma_i(l)$.

According to (\ref{algorithm01yi03}) and (\ref{algorithm01sigma07}), we are able to infer
\setlength\arraycolsep{0.5pt}
\begin{eqnarray}
y_i(l+1)-\frac{\mathbf{1}_N^T}{N}\hat{\sigma}(l+1)&=& \frac{ B_1+B_2+B_3+B_4}{[H(l:0)\mathbf{1}_N]_i+{\Psi}_i(l)N}-\nonumber \\
&&\frac{B_5 + B_6}{N([H(l:0)\mathbf{1}_N]_i+{\Psi}_i(l)N)}, \quad \quad \label{algorithm01yi04}
\end{eqnarray}
where $B_1$, $B_2$, $B_3$, $B_4$, $B_5$, $B_6$ are defined as follows
\begin{eqnarray}
B_1&=&[\big(\mathop{\Pi}\limits_{m=0}^{l}\rho(m)\big) H(l,0)\hat{\sigma}(0)]_i, \nonumber \\
B_2&=&\big(\mathop{\Pi}\limits_{m=r}^{l}\rho(m)\big)\sum_{r = 0}^{l-1}[ H(l,r)\varepsilon(r)]_i, \nonumber \\
B_3&=&\big(\mathop{\Pi}\limits_{m=r}^{l}\rho(m)\big)\sum_{r = 1}^{l}[ H(l,r){\Delta}\phi(r)]_i,\nonumber \\
B_4&=&\rho(l)[H(l)\varepsilon(l)]_i,\nonumber \\
B_5&=&\rho(l)[H(l,0)\mathbf{1}_N]_i\mathbf{1}_N^T{N}\hat{\sigma}(l),\nonumber \\
B_6&=&\rho(l)[H(l,0)\mathbf{1}_N]_i\mathbf{1}_N^T{N}\varepsilon(l). \nonumber
\end{eqnarray}

Then the follows can be deduced
\setlength\arraycolsep{2pt}
\begin{eqnarray}
&&\mathbb{E}[ \Vert y_i(l+1)-\frac{\mathbf{1}_N^T \hat{\sigma}(l+1)}{N} \Vert] \le \frac{1}{\bar{\delta}}(\mathbb{E}[\Vert B_1\Vert]+ \mathbb{E}[\Vert B_2\Vert]\nonumber \\
&& +\mathbb{E}[\Vert B_3\Vert]+\mathbb{E}[\Vert B_4\Vert])+ \frac{1}{N\bar{\delta}}(\mathbb{E}[\Vert B_5\Vert]+\mathbb{E}[\Vert B_6\Vert]). \label{Eysigma02}
\end{eqnarray}

From Assumption 6, we can obtain
\begin{eqnarray}
\mathbb{E}[\Vert B_1\Vert] &\le& N C_1\lambda^l \Vert \hat{\sigma}(0) \Vert_1 \triangleq E_1\label{E1}, \label{algorithm01E1} \\
 \mathbb{E}[\Vert B_2\Vert] &\le& \sqrt{2}N^2 C_1\sum_{r = 0}^{l-1} \lambda^{l-r}\rho(r)\check{b}(r), \nonumber
\end{eqnarray}
where $0<\mathop{\Pi}\limits_{m=r}^{l-1}\rho(m)<1$ is used. According to Lemma 1 and Assumption 6, we can get
\begin{eqnarray}
\mathbb{E}[\Vert B_2\Vert] &\le& \sqrt{2}N^2 C_1\sum_{r = 0}^{l-1} \lambda^{l-r}\rho(r)\check{b}(r)< \infty, \nonumber
\end{eqnarray}
then the constant $E_2$ is defined as the upper bound of $\mathbb{E}[|B_2|]$
\begin{eqnarray}
\mathbb{E}[\Vert B_2\Vert] \le E_2.   \label{algorithm01E2}
\end{eqnarray}

According to Assumption 2, we can know that there exists a $\hat{C_0}$, where $\hat{C_0}>0$, such that for $\forall i \in \mathcal{X}$ and $\forall t \ge 0$, the following equations hold
\begin{eqnarray}
\Vert q_i(l)\Vert &\le& \hat{C_0},  \label{xit0} \\
\Vert {\Delta}\phi(l)\Vert &=& \Vert \phi(q(l))-\phi(q(l-1))\Vert \le 2N \bar{L}_{2}\hat{C_0} \triangleq C_2, \quad \label{phiBound}
\end{eqnarray}
where $\bar{L}_2=\mathop{\text{max}}\limits_{i \in \mathcal{X}} \{L_{i_2}\}$. Then we can get $\mathbb{E}[\Vert B_3\Vert] \le N C_1 C_2 \sum_{r = 1}^{l} \lambda^{l-r}$. According to Lemma 1, $\mathop{\text{lim}}\limits_{l \to \infty} \sum_{r = 0}^{l}\lambda^{t-r}=\frac{1}{1-\lambda}$ can be obtained, so the finite sequence $\sum_{r = 1}^{l} \lambda^{l-r}< \frac{1}{1-\lambda} < \infty$ is hold. We define the up bound of $\mathbb{E}[\Vert B_3\Vert]$ by using a positive constant $E_3$
\begin{eqnarray}
\mathbb{E}[\Vert B_3\Vert] \le E_3. \label{algorithm01E3}
\end{eqnarray}

Combining the equation (\ref{algorithm01sigma04}), the following equality can be given
\setlength\arraycolsep{1pt}
\begin{eqnarray}
\rho(l)\mathbf{1}_N^T\hat{\sigma}(l) &=& \big(\mathop{\Pi}\limits_{m=0}^l\rho(m)\big) \mathbf{1}_N^T\hat{\sigma}(0) \sum_{r = 0}^{l-1} \big(\mathop{\Pi}\limits_{m=r}^l\rho(m)\big)\mathbf{1}_N^T\varepsilon(r)  \nonumber \\
&&+\sum_{r = 1}^{l} \big(\mathop{\Pi}\limits_{m=r}^l\rho(m)\big)\mathbf{1}_N^T{\Delta}\phi(r). \nonumber %\label{algorithm01sigma08E4}
\end{eqnarray}

Then the bound of $\mathbb{E}[|B_4|]$ can also be derived
\setlength\arraycolsep{2pt}
\begin{eqnarray}
\mathbb{E}[\Vert B_4\Vert] &\le&  N^2 C_1 \Vert \hat{\sigma}(0) \Vert_1+N^2\sum_{r = 0}^{l-1}\lambda^{l-r}\rho(s)\check{b}(r) \quad \quad \quad \quad \nonumber \\
&&+N^2 C_1 C_2\sum_{r = 1}^{l}\lambda^{l-s}\le E_4, \label{algorithm01sigma08E4}
\end{eqnarray}
where the last inequality is obtained by the same analysis as the obtained $E_3$, and $\lambda^{l} \le \lambda^{l-r}$ for $l \ge r$ is used. $E_4$ is the positive constant.

$\mathbb{E}[\Vert B_5\Vert]$ and $\mathbb{E}[\Vert B_6\Vert]$ can be obtained as follows according to Assumption 6
\begin{eqnarray}
\mathbb{E}[\Vert B_5\Vert] &\le& N C_1 \rho(l)\check{b}(l) \le E_5, \label{algorithm01E5} \\
\mathbb{E}[\Vert B_6\Vert] &\le& N^2 C_1 \rho(l)\check{b}(l) \le E_6, \label{algorithm01E6}
\end{eqnarray}
the analysis of $E_5$ and $E_6$ is like that of $E_3$ obtained above.

Combining formulas (\ref{Eysigma02})-(\ref{algorithm01E6}), we can obtain the upper bound of $\mathbb{E}[ \Vert y_i(l+1)-\frac{\mathbf{1}_N^T \hat{\sigma}(l+1)}{N} \Vert]$
\begin{eqnarray}
\mathbb{E}[ \Vert y_i(l+1)&-&\frac{\mathbf{1}_N^T \hat{\sigma}(l+1)}{N} \Vert] \le C_5, \label{Eysigma03}
\end{eqnarray}
where $C_5=\frac{1}{\bar{\delta}}(E_1+E_2+E_3+E_4)+ \frac{1}{N\bar{\delta}}(E_5+E_6)$.

The proof of Lemma 3 is complete.

\emph{Lemma 4.} If Assumptions 1-6 hold, the following results can be concluded
\begin{eqnarray}
\mathbb{E}[\Vert g_{i}(q_{i}(l),\sigma(l)) \Vert] &\le& M_1,\ \nonumber \\
\mathbb{E}[\Vert g_{i}(q_{i}(l),y_i(l)) \Vert] &\le& M_2, \nonumber
\end{eqnarray}
where $M_1>0, M_2>0$.

\emph{Proof: } From (\ref{phiBound}), we can know that both $\Vert \phi_i(q_i)\Vert$ and $\Vert{\sigma}(q(l))=\frac{1}{N}\sum_{i = 1}^N \phi_i(q_i)\Vert$ are bounded, which is noted as $\Vert{\sigma}(q(l))\Vert \le C_3$. According to Assumption 4, the mapping $g_{i}(q_{i}(l),{\sigma})$ is continuous for ${\sigma}(q) \in \mathbb{R}$, thus $\mathbb{E}[\Vert g_{i}(q_{i}(l),\sigma(l)) \Vert] \le M_1$ is obtained. In the next step, we need to prove $\mathbb{E}[\Vert g_{i}(q_{i}(l),y_i(l)) \Vert]$ is bounded.
\setlength\arraycolsep{1pt}
\begin{eqnarray}
\mathbb{E}[\Vert g_{i}(q_{i}(l),y_i(l)) \Vert] &\le& \bar{L}_{1}C_5+M_1\nonumber \\
&&+\bar{L}_{1}\mathbb{E}\Big[\Big\Vert \frac{\mathbf{1}_N^T \hat{\sigma}(l)}{N}-\sigma(l)\Big\Vert\Big], \quad \quad \label{gxy}
\end{eqnarray}
where $\bar{L}_{1}=\mathop{\text{max}}\limits_{i \in \mathcal{X}} \{L_{i_1}\}$.

According to (\ref{algorithm01sigma04}), we have
\setlength\arraycolsep{0.3pt}
\begin{eqnarray}
\frac{\mathbf{1}_N^T}{N}\hat{\sigma}(l) &=& \big(\mathop{\Pi}\limits_{m=0}^{l-1}\rho(m)\big) \mathbf{1}_N^T\hat{\sigma}(0)+ \sum_{r = 0}^{l-1} \big(\mathop{\Pi}\limits_{m=r}^{l-1}\rho(m)\big)\mathbf{1}_N^T\varepsilon(r) \nonumber \\
&&+\sum_{r = 1}^{l-1} \big(\mathop{\Pi}\limits_{m=r}^{l-1}\rho(m)\big)\mathbf{1}_N^T{\Delta}\phi(r) +\sigma(l)-\sigma(l-1). \ \quad \ \label{algorithm01sigma05}
\end{eqnarray}

Combining (\ref{algorithm01sigma05}), the subsequent result is attainable
\begin{eqnarray}
&&\mathbb{E}\Big[\Big\Vert \frac{\mathbf{1}_N^T}{N}\hat{\sigma}(l)-\sigma(l) \Big\Vert\Big]\nonumber \\
&\le&  N\Vert \hat{\sigma}(0) \Vert_1+\sum_{r = 0}^{l-1}\rho(r)\check{b}(r) +C_2\sum_{r = 1}^{l-1}\rho(r)^{l-r}+C_3,
 \label{HatAndsigma01}
\end{eqnarray}
from the same analysis of the obtained (\ref{algorithm01E5}) and (\ref{algorithm01E3}), $\mathbb{E}\Big[\Big\Vert \frac{\mathbf{1}_N^T}{N}\hat{\sigma}(l)-\sigma(l) \Big\Vert\Big] < \infty$ can be obtained. Thus, we can get $\mathbb{E}\Big[\Big\Vert \frac{\mathbf{1}_N^T}{N}\hat{\sigma}(l)-\sigma(l) \Big\Vert\Big] \le N_1$. Then we can be obtained $\mathbb{E}[\Vert g_{i}(q_{i}(l),y_i(l)) \Vert]\le  \bar{L}_{1}C_5+\bar{L}_{1}N_1+M_1 \triangleq M_2$. The proof of Lemma 4 is complete.

\emph{Lemma 5.} Let $\bar{\beta}=\mathop{\text{max}}\limits_{j \in \mathcal{X}} \{\beta_j\}$. If $0<\bar{\beta}<\frac{\sqrt{2} }{2}$ and Assumptions 1-6 hold, then for any $i \in \mathcal{X}$, the following result is obtained:

\noindent(1) $\mathbb{E}[\Vert q_i(l+1)-q_i(l) \Vert] \le M_2\sum_{r = 1}^l\bar{\beta}^{l-r}\bar{\mu}(r)$, $\forall l \in \mathbb{N}_+$;

\noindent(2)$\mathop{\text{lim}}\limits_{l \to \infty} \mathbb{E}[\Vert q_i(l+1)-q_i(l) \Vert]=0$;

\noindent(3) $\sum_{l = 1}^{\infty}\mathbb{E}[\Vert q_i(l+1)-q_i(l) \Vert^2] <\infty$.

\emph{Proof: } \noindent(1) According to the equation (\ref{algorithm01xi}) and the non-expansive property of the projection operator, we can derive that $\mathbb{E}[\Vert q_i(l+1)-q_i(l) \Vert] \le   M_2 \sum_{r = 1}^{l} \bar{\beta}^{l-r}\bar{\mu}(r)$, where $\Vert q_i(1)-q_i(0) \Vert=0$ is used in this inequality.

\noindent(2) According to Assumption 5, $\mathop{\text{lim}}\limits_{l \to \infty} \mu_i(l)=0$ can be obtained by $\sum_{l = 1}^{\infty}(\mu_i(l))^2 <\infty$. Thus $\mathop{\text{lim}}\limits_{l \to \infty} \bar{\mu}(l)=0$, then $\mathop{\text{lim}}\limits_{l \to \infty} \sum_{r = 1}^{l} \bar{\beta}^{l-r}\bar{\mu}(r)=0$ can be obtained by Lemma 1. Therefore, $\mathop{\text{lim}}\limits_{l \to \infty} \mathbb{E}[\Vert q_i(l+1)-q_i(l) \Vert]=0$.

\noindent(3) According to the first item (1) of Lemma 5, we have $\mathbb{E}[\Vert q_i(l+1)-q_i(l) \Vert^2]  \le 2M_2^2 \sum_{r = 1}^{l} (2\bar{\beta}^2)^{l-r}\bar{\mu}^2(r)$.

\emph{Remark 6.} Note that the momentum term $\beta_{i}(q_{i}(l)-q_{i}(l-1))$ plays an important role in ensuring the boundness of $\mathbb{E}[\Vert q_i(l+1)-q_i(l) \Vert^2]$. This is also the basis for the following proving that the strategies of players can almost surely achieve NE.

Because $0<\bar{\beta}<\frac{\sqrt{2} }{2}$, then $0<2\bar{\beta}^2<1$ can be obtained. $\sum_{l = 1}^{\infty}\sum_{r = 1}^{l} (2\bar{\beta}^2)^{l-r}\bar{\mu}^2(r)<\infty$ can be obtained by $\sum_{l = 1}^{\infty}( \mu_i(l) )^2 <\infty$ and Lemma 1. Thus, $\sum_{l = 1}^{\infty}\mathbb{E}[\Vert q_i(l+1)-q_i(l) \Vert^2] <\infty$ is proved.

\emph{Lemma 6.} If Assumptions 1-6 hold, the following results can be obtained:

\noindent(1) $\mathop{\text{lim}}\limits_{l \to \infty} \mathbb{E}[\Vert y_i(l+1)- \frac{\mathbf{1}_N^T \hat{\sigma}(l+1)}{N} \Vert]=0$;

\noindent(2) $\sum_{l = 1}^{\infty} \bar{\mu}(l) \mathbb{E}[ \Vert y_i(l+1)- \frac{\mathbf{1}_N^T \hat{\sigma}(l+1)}{N} \Vert] < \infty$.

\emph{Proof: } \noindent(1) From (\ref{Eysigma02}), we can get
\setlength\arraycolsep{2pt}\
\begin{eqnarray}
&&\mathbb{E}[ \Vert y_i(l+1)-\frac{\mathbf{1}_N^T \hat{\sigma}(l+1)}{N} \Vert]  \nonumber \\
&\le& \frac{1}{\bar{\delta}}\Big(N C_1\lambda^t \Vert \hat{\sigma}(0) \Vert_1 +\sqrt{2}N^2 C_1 \sum_{r = 0}^{l-1}\big(\mathop{\Pi}\limits_{m=r}^{l}\rho(m)\big) \lambda^{l-r}\check{b}(r)\nonumber \\
&&+2N C_1\bar{L}_{2}\sum_{r = 1}^{l}\big(\mathop{\Pi}\limits_{m=r}^{l}\rho(m)\big)\lambda^{l-r}\Vert q(r)-q(r-1) \Vert \quad \quad  \nonumber \\
&&+N C_1 \rho(l)\check{b}(l)\Big) \nonumber \\
&&+\frac{1}{N\bar{\delta}}\Big(N^2 C_1\lambda^l \Vert \hat{\sigma}(0) \Vert_1+N^2 C_1\lambda^l\rho(l)\check{b}(l)\nonumber \\
&&+C_1\bar{L}_{2}\sum_{r = 1}^{l}\lambda^{l-r}\big(\mathop{\Pi}\limits_{m=r}^{l}\rho(m)\big)\Vert q(r)-q(r-1) \Vert\nonumber \\
&&+N^2 C_1\sum_{r = 0}^{l-1}\big(\mathop{\Pi}\limits_{m=r}^{l}\rho(m)\big) \lambda^{l-r}\check{b}(r)\Big). \label{Eysigma0Lemma0601}
\end{eqnarray}

Because $\lambda \in (0,1)$, then $\mathop{\text{lim}}\limits_{l \to \infty}N C_1\lambda^l \Vert \hat{\sigma}(0) \Vert_1=0$. And $\mathop{\text{lim}}\limits_{l \to \infty} N C_1 \rho(l)\check{b}(l)=0$, $\mathop{\text{lim}}\limits_{l \to \infty} N C_1\lambda^l \rho(l)\check{b}(l)=0$ can be obtained by Assumption 6. In addition, according to Lemma 1 and Assumption 6, we have $\mathop{\text{lim}}\limits_{l \to \infty}(\sqrt{2}N^2 C_1 \sum_{r = 0}^{l-1}\big(\mathop{\Pi}\limits_{m=r}^{l}\rho(m)\big) \lambda^{l-r}\check{b}(r))\le0$. According to Lemma 5 and Lemma 1, there holds that $ \mathop{\text{lim}}\limits_{l \to \infty}(2N C_1\bar{L}_{2}\sum_{r =1}^{l}\big(\mathop{\Pi}\limits_{m=r}^{l}\rho(m)\big)\lambda^{l-r}\Vert q(r)-q(r-1) \Vert) \le0$. And one gets that $\mathop{\text{lim}}\limits_{l \to \infty}(N^2 C_1\sum_{r = 0}^{l-1}\big(\mathop{\Pi}\limits_{m=r}^{l}\rho(m)\big) \lambda^{l-r}\check{b}(r))\le0$ from Assumption 6. In addition, $\mathop{\text{lim}}\limits_{l \to \infty}(C_1\bar{L}_{2}\sum_{r =1}^{l}\lambda^{l-r}\big(\mathop{\Pi}\limits_{m=r}^{l}\rho(m)\big)\Vert q(r)-q(r-1) \Vert) \le0$ can also be obtained from Lemma 5, Assumption 6 and Lemma 1. Then $\mathop{\text{lim}}\limits_{l \to \infty} \mathbb{E}\Big[\Big\Vert y_i(l+1)- \frac{\mathbf{1}_N^T \hat{\sigma}(l+1)}{N} \Big\Vert\Big]=0$ is proved.

\noindent(2) According to (\ref{Eysigma0Lemma0601}), the following equation can be obtained
\setlength\arraycolsep{2pt}
\begin{eqnarray}
&&\sum_{l = 1}^{\infty} \bar{\mu}(l) \mathbb{E}[ \Vert y_i(l+1)- \frac{\mathbf{1}_N^T \hat{\sigma}(l+1)}{N} \Vert]  \nonumber \\
&\le&\frac{1}{\bar{\delta}}\Big(\sum_{l = 1}^{\infty} \bar{\mu}(l)N C_1\lambda^l \Vert \hat{\sigma}(0) \Vert_1\nonumber \\
&&+\sum_{l = 1}^{\infty} \bar{\mu}(l)\sqrt{2}N^2 C_1 \sum_{r = 0}^{l-1}\big(\mathop{\Pi}\limits_{m=r}^{l}\rho(m)\big) \lambda^{l-r}\check{b}(r) \quad \quad \quad\nonumber \\
&&+\sum_{l = 1}^{\infty} \bar{\mu}(l)2N C_1C_2\sum_{r = 1}^{l}\big(\mathop{\Pi}\limits_{m=r}^{l}\rho(m)\big)\lambda^{l-r}\nonumber \\
&&+\sum_{l = 1}^{\infty} \bar{\mu}(l)N C_1\rho(l)\check{b}(l)\Big)\nonumber \\
&&+ \frac{1}{N\delta}\Big(\sum_{l = 1}^{\infty} \bar{\mu}(l)N^2 C_1\lambda^l \Vert \hat{\sigma}(0) \Vert_1+\sum_{l = 1}^{\infty} \bar{\mu}(l)N^2 C_1\lambda^l\rho(l)\check{b}(l)\nonumber \\
&&+\sum_{l = 1}^{\infty} \bar{\mu}(l)C_1C_2\sum_{r = 1}^{l}\lambda^{l-r}\big(\mathop{\Pi}\limits_{m=r}^{l}\rho(m)\big)\nonumber \\
&&+\sum_{l = 1}^{\infty} \bar{\mu}(l)N^2 C_1\sum_{r = 0}^{l-1}\big(\mathop{\Pi}\limits_{m=r}^{l}\rho(m)\big) \lambda^{l-r}\check{b}(r)\Big). \label{Eysigma0Lemma0602}
\end{eqnarray}

And the following inequality can be inferred: $\sum_{l = 1}^{\infty} \bar{\mu}(l)N C_1\lambda^l \Vert \hat{\sigma}(0) \Vert_1 < \infty$, where Assumption 5 and Lemma 1 are used. In addition, $\sum_{l = 1}^{\infty} \bar{\mu}(l)\sqrt{2}N^2 C_1 \sum_{r = 0}^{l-1}\big(\mathop{\Pi}\limits_{m=r}^{l}\rho(m)\big) \lambda^{l-r}\check{b}(r) <\infty$, where Assumption 5, Assumption 6 and Lemma 1 are used. Then according to the same analysis as above, the following inequalities can also be achieved:
\setlength\arraycolsep{1pt}
\begin{eqnarray}
\sum_{l = 1}^{\infty} \bar{\mu}(l)N C_1\rho(l)\check{b}(l) &<& \infty,\nonumber \\
\sum_{l = 1}^{\infty} \bar{\mu}(l)2N C_1C_2\sum_{r = 1}^{l}\big(\mathop{\Pi}\limits_{m=r}^{l}\rho(m)\big)\lambda^{l-r}
&<& \infty,\nonumber \\
\sum_{l = 1}^{\infty} \bar{\mu}(l)N^2 C_1\lambda^l \Vert \hat{\sigma}(0) \Vert_1 &<& \infty,  \nonumber \\
\sum_{l = 1}^{\infty} \bar{\mu}(l)N^2 C_1\lambda^l\rho(l)\check{b}(l)&<& \infty, \nonumber \\
\sum_{l = 1}^{\infty} \bar{\mu}(l)C_1C_2\sum_{r = 1}^{l}\lambda^{l-r}\big(\mathop{\Pi}\limits_{m=r}^{l}\rho(m)\big) &<& \infty, \nonumber
\end{eqnarray}
\setlength\arraycolsep{1pt}
\begin{eqnarray}
\sum_{l = 1}^{\infty} \bar{\mu}(l)N^2 C_1\sum_{r = 0}^{l-1}\big(\mathop{\Pi}\limits_{m=r}^{l}\rho(m)\big) \lambda^{l-r}\check{b}(r)&<& \infty. \nonumber
\end{eqnarray}

Therefore, $\sum_{l = 1}^{\infty} \bar{\mu}(l) \mathbb{E}\Big[ \Big \Vert y_i(l+1)- \frac{\mathbf{1}_N^T \hat{\sigma}(l+1)}{N} \Big \Vert \Big] < \infty$ is obtained. The proof of Lemma 6 is complete.

The final convergence result of the algorithm is presented in Theorem 1 below.

\emph{Theorem 1.} Take into account the aggregative game (\ref{obiectFunc1}) involving $N$ players across a time-varying digraph $\mathcal{T}(l)$. Assume that Assumptions 1-6 hold true. If $0<\beta_i<\frac{\sqrt{2} }{2},\ \forall i \in \mathcal{X}$, and every player updates its strategy according to the algorithms (\ref{algorithm01si})-(\ref{algorithm01xi}), then the actions of all players will converge almost surely to the NE $q^*$.

\emph{Proof: } Denote $\sigma(q^*)$ as $\sigma^*$. According to (\ref{algorithm01xi}) the following result can be inferred
\setlength\arraycolsep{2pt}
\begin{eqnarray}
&&\mathbb{E}[ \Vert q_{i}(l+1)- q_{i}^* \Vert^2] \nonumber \\
&\le& \mathbb{E}[\Vert  q_{i}(l)-q_{i}^*(l)-\mu_{i}(l)(g_{i}(q_{i}(l),y_i(l)) \quad  \quad \quad \quad \quad \quad \nonumber \\
&&-g_{i}(q^*_{i}(l),\sigma^*))\Vert^2] +\beta_{i}^2\mathbb{E}[ \Vert q_{i}(l)-q_{i}(l-1) \Vert^2]\nonumber \\
&&+2\beta_{i}\mathbb{E}[(q_{i}(l)-q_{i}(t-1))(q_{i}(l)-q_{i}^*(l))]\nonumber \\
&&-2\beta_{i}\mu_{i}(l)\mathbb{E}[(q_{i}(l)-q_{i}(l-1)) (g_{i}(q_{i}(l),y_i(l))\nonumber \\
&&-g_{i}(q^*_{i}(l),\sigma^*))].  \quad \quad \label{algorithm01xi0Theorem01}
\end{eqnarray}

For the last term of (\ref{algorithm01xi0Theorem01}), we can obtain
\setlength\arraycolsep{2pt}
\begin{eqnarray}
&&-2\beta_{i}\mu_{i}(l)\mathbb{E}[(q_{i}(l)-q_{i}(l-1))(g_{i}(q_{i}(l),y_i(l))\nonumber \\
&& -g_{i}(q^*_{i}(l),\sigma^*))]\nonumber \\
&\le&  4M_2^2 \bar{\mu}^2(l)+\beta^2_{i} \vert q_{i}(l)-q_{i}(l-1) \vert^2, \quad \quad \quad \quad \quad \quad
\label{algorithm01xi0Theorem02}
\end{eqnarray}
where Lemma 4 is used in the last inequality.

Adding both left side and right side of the inequality (\ref{algorithm01xi0Theorem01}) from $i = 1$ to $N$ yields
\setlength\arraycolsep{2pt}
\begin{eqnarray}
&&\mathbb{E}[ \Vert q(l+1)- q^* \Vert^2] \nonumber \\
&\le&\sum_{i = 1}^{N}\mathbb{E}[\Vert  q_{i}(l)-q_{i}^*(l)-\mu_{i}(l)(g_{i}(q_{i}(l),y_i(l)) \nonumber \\
&& -g_{i}(q^*_{i}(l),\sigma^*))\Vert^2]+2\bar{\beta}^2\mathbb{E}[ \Vert q(l)-q(l-1) \Vert^2]\nonumber \\
&&+2\mathbb{E}[(q(l)-q(l-1))^T\Theta(q(l)-q^*(l))]\nonumber \\
&&+4NM_2^2 \bar{\mu}^2(l), \label{algorithm01xi0Theorem03}
\end{eqnarray}
where $\Theta=$diag$\{\beta_{1},\cdots,\beta_{N}\}$, diag$\{\beta_i\}$ denotes a $N \times N$ diagonal matrix. Furthermore, the following inequality can be provided
\begin{eqnarray}
&&2\mathbb{E}[(q(l)-q(l-1))^T\Theta(q(l)-q^*)] \nonumber \\
&\le& \mathbb{E}[\Vert q(l)-q^* \Vert_{\Theta}^2] +\bar{\beta}\mathbb{E}[\Vert q(l)-q(l-1) \Vert^2] \nonumber \\
&&-\mathbb{E}[\Vert q(l-1)-q^* \Vert_{\Theta}^2], \nonumber
\end{eqnarray}
where $\Vert q(l-1)-q^*(l) \Vert_{\Theta}=(q(l-1)-q^*)^T\Theta(q(l-1)-q^*)$.

Thus, the following inequality can be deduced
\setlength\arraycolsep{2pt}
\begin{eqnarray}
&&\Vert q_{i}(l)-q_{i}^*-\mu_{i}(l)(g_{i}(q_{i}(l),y_i(l))-g_{i}(q^*_{i},\sigma^*))\Vert^2\nonumber \\
&\le&\Vert q_{i}(l)-q_{i}^*\vert^2+4M_2^2\bar{\mu}(l)+4C_0\bar{L}_{2}\bar{\mu}(l)\Vert y_i(l)-\sigma(l)\Vert \nonumber \\
&&-2(\mu_{i}(l)-\underline{\mu}(l))(q_{i}(l)-q_{i}^*)(g_{i}(q_{i}(l),\sigma(l))\nonumber \\
&&-g_{i}(q^*_{i},\sigma^*))-2\underline{\mu}(l)(q_{i}(l)-q_{i}^*)(g_{i}(q_{i}(l),\sigma(l))\nonumber \\
&&-g_{i}(q^*_{i},\sigma^*)), \quad  \quad   \quad  \quad \quad  \quad
\label{algorithm01xi0Theoremmiddle}
\end{eqnarray}
where Assumption 4 and Lemma 4 are used.

Because $\Vert y_i(l)-\sigma(l)\Vert \le \Vert y_i(l)- \frac{\mathbf{1}_N^T \hat{\sigma}(l)}{N}\Vert+\Vert \frac{\mathbf{1}_N^T \hat{\sigma}(l)}{N}-\sigma(l)\Vert$, one can get that
\setlength\arraycolsep{1pt}
\begin{eqnarray}
&&\Vert q_{i}(l)-q_{i}^*-\mu_{i}(l)(g_{i}(q_{i}(l),y_i(l))-g_{i}(q^*_{i},\sigma^*))\Vert^2\nonumber \\
&\le& \Vert q_{i}(l)-q_{i}^*\Vert^2+4M_2^2\bar{\mu}(l)+ 4C_0\bar{L}_{2}\bar{\mu}(l)\cdot\nonumber \\
&&\Vert y_i(l)- \frac{\mathbf{1}_N^T \hat{\sigma}(l)}{N}\Vert +4C_0\bar{L}_{2}\bar{\mu}(l)\Vert \frac{\mathbf{1}_N^T \hat{\sigma}(l)}{N}-\sigma(l)\Vert\nonumber \\
&&+8 C_0 M_1(\bar{\mu}(l)-\underline{\mu}(l)) \nonumber \\
&&-2\underline{\mu}(l) (q_{i}(l)-q_{i}^*)(g_{i}(q_{i}(l),\sigma(t))\nonumber \\
&&-g_{i}(q^*_{i},\sigma^*)). \quad \quad \label{algorithm01xi0Theoremmiddle02}
\end{eqnarray}

Note that $g_{i}(q_{i}(l),\sigma(l))$ is equal to the value of $\nabla_i J_i(q)$, then adding both right side and left side of the inequality (\ref{algorithm01xi0Theoremmiddle02}) from $i = 1$ to $N$ results in
\setlength\arraycolsep{1pt}
\begin{eqnarray}
&&\sum_{i = 1}^{N}\Vert q_{i}(l)-q_{i}^*-\mu_{i}(l)(g_{i}(q_{i}(l),y_i(l))-g_{i}(q^*_{i},\sigma^*))\Vert^2\nonumber \\
&\le&  \Vert q(l)-q^*\Vert^2+4NM_2^2\bar{\mu}(l)+ 4C_0\bar{L}_{2}\bar{\mu}(l)\Vert y(l)- \nonumber \\
&& \mathbf{1}_N\frac{\mathbf{1}_N^T \hat{\sigma}(l)}{N}\Vert +4NC_0\bar{L}_{2}\bar{\mu}(l)\Vert \frac{\mathbf{1}_N^T \hat{\sigma}(l)}{N}-\sigma(l)\Vert+\nonumber \\
&&8 NC_0 M_1(\bar{\mu}(l)-\underline{\mu}(l))- \nonumber \\
&& 2\underline{\mu}(l)(q(l)-q^*)^T  (\nabla J(q)-\nabla J(q^*)). \label{algorithm01xi0Theoremmiddle03}
\end{eqnarray}

By the fact that $\bar{\beta}^2<\bar{\beta}$, and combining (\ref{algorithm01xi0Theorem03}) and (\ref{algorithm01xi0Theoremmiddle03}), the following inequality can be got
\setlength\arraycolsep{2pt}
\begin{eqnarray}
&&\mathbb{E}[ \Vert q(l+1)- q^* \Vert^2]-\mathbb{E}[\Vert q(l)-q^*\Vert^2] \nonumber \\
&\le&3\bar{\beta}\mathbb{E}[ \Vert q(l)-q(l-1) \Vert^2]+\mathbb{E}[\Vert q(l)-q^* \Vert_{\Theta}^2]\nonumber \\
&&-\mathbb{E}[\Vert q(l-1)-q^* \Vert_{\Theta}^2]+F(l) \nonumber \\
&&
-2\underline{\mu}(l)(q(l)-q^*)^T(\nabla J(q)-\nabla J(q^*)). \quad \quad \quad \quad\label{algorithm01xi0Theorem05}
\end{eqnarray}

Therefore,
\setlength\arraycolsep{2pt}
\begin{eqnarray}
&&2\sum_{l = 1}^{\infty}\underline{\mu}(l)(q(l)-q^*)^T(\nabla J(q)-\nabla J(q^*)) \nonumber \\
&&\le 3\bar{\beta} \sum_{l = 1}^{\infty}\mathbb{E}[ \Vert q(l)-q(l-1) \Vert^2]+\mathbb{E}[\Vert q(l)-q^* \Vert_{\Theta}^2]\nonumber \\
&&-\mathbb{E}[ \Vert q(l+1)- q^* \Vert^2] + \mathbb{E}[ \Vert q(1)- q^* \Vert^2]\nonumber \\
&&-\mathbb{E}[\Vert q(0)-q^*(l) \Vert_{\Theta}^2]+\sum_{l = 1}^{\infty}F(l), \label{algorithm01xi0Theorem06}
\end{eqnarray}
where $F(l)=8NM_2^2\bar{\mu}(l)+ 4C_0\bar{L}_{2}\bar{\mu}(l)\mathbb{E}[\Vert y(l)- \mathbf{1}_N\frac{\mathbf{1}_N^T \hat{\sigma}(l)}{N}\Vert] +4NC_0\bar{L}_{2}\bar{\mu}(l)\mathbb{E}[\Vert \frac{\mathbf{1}_N^T \hat{\sigma}(l)}{N}-\sigma(l)\Vert]+8 NC_0 M_1(\bar{\mu}(l)-\underline{\mu}(l))$.

Combining Assumption 5, Assumption 6 and Lemmas 3-6, $\sum_{l = 1}^{\infty}F(l)<\infty$ can be obtained. From Lemma 5, $\mathop{\text{lim}}\limits_{l \to \infty} \mathbb{E}[\Vert q_i(l+1)-q_i(l) \Vert]=0$, then the sequence $\{q(l)\}$ is convergent. Therefore, $\mathop{\text{lim}}\limits_{l \to \infty} (\mathbb{E}[\Vert q(l)- q^* \Vert_{\Theta}^2]-\mathbb{E}[ \Vert q(l+1)- q^* \Vert^2])\le0$ by the fact that $\beta_i \in (0,1)$. Then combining Lemma 5 and the above analysis, one has $\sum_{l = 1}^{\infty}\underline{\mu}(l)(q(l)-q^*)^T(\nabla J(q)-\nabla J(q^*)) < \infty$. Since $\sum_{l = 1}^{\infty}( \mu_i(l) ) = \infty$, then $\mathop{\text{lim}}\limits_{l \to \infty}(q(l)-q^*)^T(\nabla J(q)-\nabla J(q^*))=0$ can be obtained by using Assumption 3. Then the proof of Theorem 1 is complete.

\section{Differential Privacy Analysis}

The information transmitted by players in our paper is the estimation of the aggregate function values, i.e., $\hat{\sigma}(l)=[\hat{\sigma}_1(l),\cdots,\hat{\sigma}_N(l)]^T$. Thus, the sensitivity of the distributed NE seeking algorithms (\ref{algorithm01si})-(\ref{algorithm01xi}) to problem (\ref{obiectFunc1}) is defined as follows:

\emph{Definition 4.} At each iteration $l$, $\widetilde{H}$ and $\widetilde{H}'$ are any two adjacent distributed NE seeking problems, for any initial state $\tau(0)$, the sensitivity of the proposed algorithm is
\begin{eqnarray}
\Delta(l) &\triangleq&\mathop{\text{sup}}\limits_{\mathcal{O}_s \in \mathbb{O}} \Big \{ \mathop{\text{sup}}\limits_{\tau(l) \in \mathcal{A}_{\widetilde{H},\mathcal{O},\tau(0)}[l],\tau '(l) \in \mathcal{A}_{\widetilde{H}',\mathcal{O},\tau(0)}[l]} \nonumber \\
&& \Vert \hat{\sigma}(l)-\hat{\sigma}'(l) \Vert_1 \Big \}. \label{Definition4SS}
\end{eqnarray}

Before giving the final theorem of differential privacy, the following lemma needs to be given for the first step.

\emph{Lemma 7.} In the distributed NE seeking algorithms (\ref{algorithm01si})-(\ref{algorithm01xi}), if a Laplace noise $\varepsilon_i(l)$ with parameter $b_i(l)$, where $\hat{b}(l) \le b_i(l) \le \check{b}(l)$, $\hat{b}(l)=\mathop{\text{min}}\limits_{i \in \mathcal{X}} \{b_i(l)\}$, is added to the shared messages $\hat{\sigma}_i(l)$ for each player such that $\sum_{l = 1}^{T}\frac{\Delta(l)}{\hat{b}(l)}\le\bar{\epsilon}$ then the iterative distributed algorithms (\ref{algorithm01si})-(\ref{algorithm01xi}) is $\epsilon$-differentially private with the cumulative privacy budget for iterations from $l = 0$ to $l = T$ less than $\bar{\epsilon}$.

\emph{Proof:} The lemma can be obtained following the same line of proof of Theorem 3 in \cite{YEDP2021TAC}.

\emph{Theorem 2.} If Assumptions 1-6 hold, and all elements of $\varepsilon_i(l)$ are independent Laplace noise with distribution Lap($b_i(l)$). Then for any finite number of iterations $T$, the designed algorithm is $\epsilon$-differential privacy with the cumulative privacy budget bounded by $\epsilon = \sum_{l = 1}^{T}\frac{\Delta(l)}{\hat{b}(l)}\le\bar{\epsilon}$, and the cumulative privacy budget is summable even though $T \rightarrow \infty$ when $\frac{\rho(l)}{\hat{b}(l)}$ is summable.

\emph{Proof: }Before obtaining the final cumulative privacy budget, we give some preparatory notes. Two distributed NE seeking problems $\widetilde{H}=\{U,\mathbb{J},\mathcal{T}(l)\}$ and $\widetilde{H}'=\{U',\mathbb{J'},\mathcal{T'}(l)\}$ are adjacent. The observation $\mathcal{O}$ and initial states $\tau(0)$ are the same for the two adjacent NE seeking problems. Since for $\widetilde{H}$ and $\widetilde{H}'$, there is only one cost function that is different, we represent this different cost function as the $i$-th one, i.e., $J_i(\cdot)$, without loss of generality. In addition, the sensitivity of the designed algorithm is determined as $\Vert \hat{\sigma}(l)-\hat{\sigma}'(l) \Vert_1$. Due to the initial conditions, cost functions, and observations of $\widetilde{H}$ and $\widetilde{H}'$ are identical for $j \neq i$, we have
$\hat{\sigma}_{j}(l)=\hat{\sigma}_{j}^{'}(l)$ for all $j = i$ and $l$. Therefore, $\Vert \hat{\sigma}(l)-\hat{\sigma}'(l) \Vert_1 = \Vert \hat{\sigma}_{i}(l)-\hat{\sigma}_{i}^{'}(l) \Vert_1$ can be obtained for all $l$. Note that $\hat{\sigma}_i(l) + \varepsilon_i(l)=\hat{\sigma} '_i(l) + \varepsilon '_i(l)$ is used in our paper, then $z_i(l+1)=z'_i(l+1)$ can also been obtained.

Then, we begin to derive the sensitivity. According to Definition 4, the update rule (\ref{algorithm01sigma}), the sensitivity (\ref{Definition4SS}) of the algorithm, and combining $\Vert {\phi}_i(q_1)-{\phi}_i(q_2)\Vert \le L_{i_2} |q_1-q_2|$ in Assumption 4, then we have
\begin{eqnarray}
&&\Vert \hat{\sigma}_i(l)-\hat{\sigma} ' _i(l) \Vert  \nonumber \\
&&= \Vert \phi_i(q_i(l))-\phi_i(q_i(l-1))-(\phi ' _i(q_i(l))-\phi '_i(q_i(l-1))) \Vert \nonumber \\
&&\le \bar{L}_2(\Vert q_i(l)-q_i(l-1)\Vert+\Vert q '_i(l)-q ' _i(l-1) \Vert). \nonumber
\end{eqnarray}

According to Definition 1, the actions of players under two adjacent distributed NE seeking problems $\widetilde{H}$ and $\widetilde{H}'$ converge to the same NE. Thus, $\Vert q_i(l)- q _i(l-1)\Vert=0$ will hold when $l$ is large enough. And the ensured almost sure convergence in Lemma 5 and Theorem 1 also means that  there exists a constant $M_3$ such that the sequence $\Vert q_i(l)-q_i(l-1)\Vert+\Vert q ' _i(l)-q ' _i(l-1)\Vert$ are upper bounded by the sequence $\bar{L}_2M_3\rho(l)$. Thus $\Delta(l)\le \bar{L}_2M_3\rho(l)$, $\epsilon \le \sum_{l = 1}^{T}\frac{\bar{L}_2M_3\rho(l)}{\hat{b}(l)}$. Then, according to Lemma 6 of the reference \cite{WYQ2024TAC02}, $\epsilon$ will be finite even when
$l$ tends to infinity if $\sum_{l=0}^{\infty}\frac{\rho(l)}{\hat{b}(l)}<\infty$. Theorem 2 has been proved.

\emph{Remark 7.} From Theorem 1, one has $\mathbb{E}[ \Vert q(l+1)- q^* \Vert^2]\le\sum_{i = 1}^{N}\mathbb{E}[\vert  q_{i}(l)-q_{i}^*(l)-\mu_{i}(l)(g_{i}(q_{i}(l),y_i(l))-g_{i}(q^*_{i}(l),\sigma^*))\vert^2]+3\bar{\beta}^2\mathbb{E}[ \Vert q(l)-q(l-1) \Vert^2]+\mathbb{E}[\Vert q(l)-q^* \Vert_{\Theta}^2] -\mathbb{E}[\Vert q(l-1)-q^* \Vert_{\Theta}^2]+4NM_2^2 \bar{\mu}^2(l)$, and $\Vert q_{i}(l)-q_{i}^*-\mu_{i}(l)(g_{i}(q_{i}(l),y_i(l))-g_{i}(q^*_{i},\sigma^*))\Vert^2= \Vert q_{i}(l)-q_{i}^*\Vert^2+\mu_{i}^2(l)(g_{i}(q_{i}(l),y_i(l))-g_{i}(q^*_{i},\sigma^*))^2 -2\mu_{i}(l)(q_{i}(l)-q_{i}^*)^T (g_{i}(q_{i}(l),y_i(l))-g_{i}(q^*_{i},\sigma^*))$. In addition, $-2\mu_{i}(l)(q_{i}(l)-q_{i}^*)^T(g_{i}(q_{i}(l),y_i(l))-g_{i}(q^*_{i},\sigma^*))
\le\frac{\underline{\mu}^2(l)}{\rho^2(l)}\vert q_{i}(l)-q_{i}^* \Vert^2+\rho^2(l)\bar{L}_{1}\Vert y_i(l)-\sigma(l) \Vert^2-2\underline{\mu}(l)(q_{i}(l)-q_{i}^*)^T(g_{i}(q_{i}(l),\sigma(l))-g_{i}(q^*_{i},\sigma^*))$ and $\Vert y_i(l)-\sigma(l) \Vert^2 \le 2\vert y_i(l)-\frac{\mathbf{1}_N^T \hat{\sigma}(l+1)}{N}\Vert^2+2\Vert\frac{\mathbf{1}_N^T}{N}\hat{\sigma}(l)-\sigma(l)\Vert^2
\le2C_5^2+2N_1^2$ can be obtained. Thus, the following inequality can be derived: $\mathbb{E}[ \Vert q(l+1)- q^* \Vert^2] \le \big(1+\frac{\underline{\mu}^2(l)}{\rho^2(l)}\big)\mathbb{E}[\Vert q(l)-q^* \Vert^2]+2\rho^2(l)\bar{L}_{1}N(C_5^2+N_1^2) +3\bar{\beta}^2\mathbb{E}[ \Vert q(l)-q(l-1) \Vert^2]+4NM_2^2 \bar{\mu}^2(l)+\mathbb{E}[\Vert q(l)-q^* \Vert_{\Theta}^2] -\mathbb{E}[\Vert q(l-1)-q^*\Vert_{\Theta}^2]-2\underline{\mu}(l)(q(l)-q^*)^T(g_{i}(q(l),\mathbf{1}_N\sigma(l))-g(q^*,\mathbf{1}_N\sigma^*))$. Then combine the results above and the similar analysis in Remark 9 of \cite{WYQ2024TAC01}, we can obtain that $\Vert q(l+1)- q^* \Vert$ decreases to zero with a rate of $\mathcal{O}(\frac{\underline{\mu}(l)}{\rho(l)})$.

\emph{Remark 8.} It's worth noting that our algorithm incurs a trade-off in terms of convergence rate. From Remark 7, the convergence rate is $\mathcal{O}(\frac{\underline{\mu}(l)}{\rho(l)})$, which means the decreasing speed to zero reduces with an increase in the decreasing speed of $\rho(l)$. From Theorem 2, we can see that, in order to reduce $\epsilon$ to enhance privacy, a faster increasing $\hat{b}(l)$ needs to be employed, which requires $\rho(l)$ to decrease faster according to Assumption 6. Then, a stronger privacy preservation needs to sacrifice faster convergence rate can be concluded.

\begin{figure}[!t]
\centering
\includegraphics[width=2.3in]{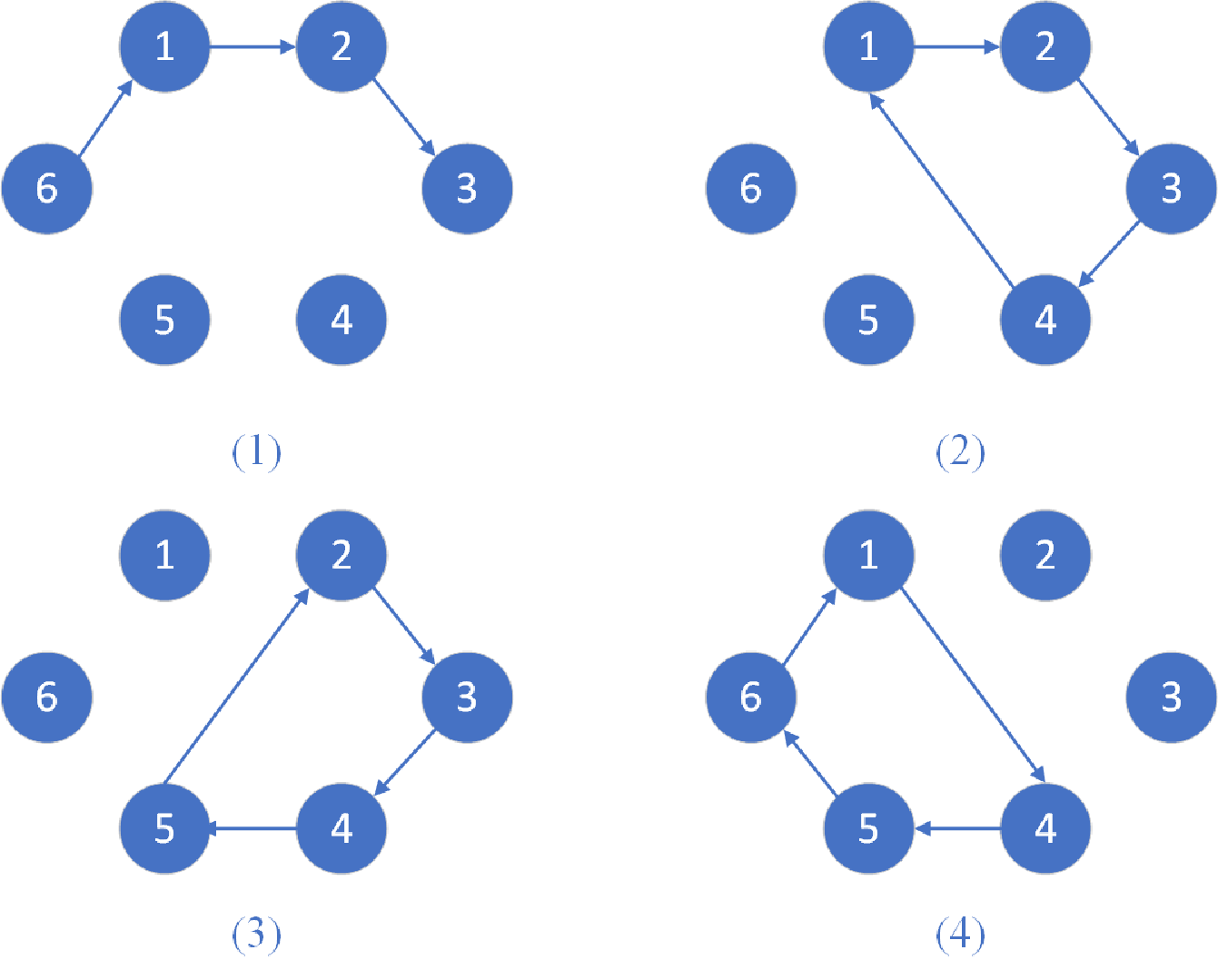}
\caption{Time-varying communication graphs with six players.} \label{fig1}
\end{figure}

\begin{figure}[t]
\centering  %图片全局居中
\subfigure[The actions of the players.]{
\label{Fig.sub.1}
\includegraphics[width=1.6in]{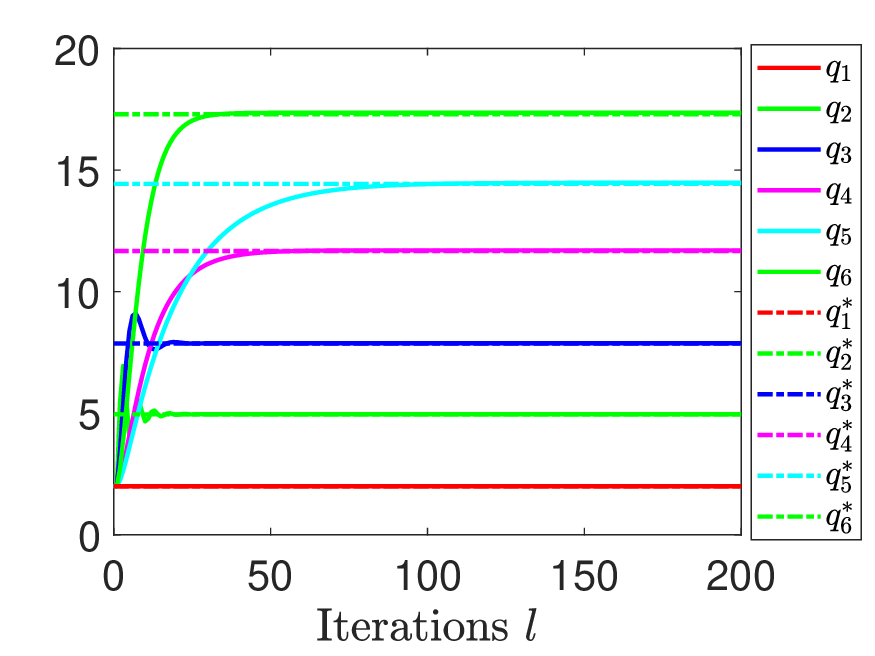}}
\subfigure[Privacy budget.]{
\label{Fig.sub.2}
\includegraphics[width=1.6in]{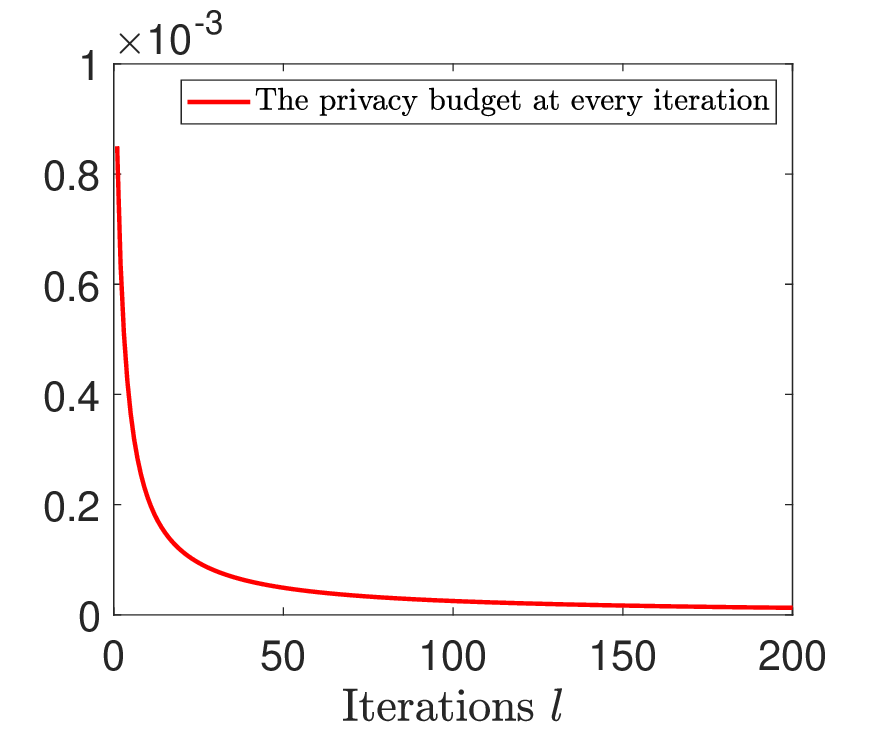}}
\caption{The actions of the players and privacy budget at each iteration under the algorithms (\ref{algorithm01si})-(\ref{algorithm01xi}).}
\label{size01}
\end{figure}

\section{Simulation}
\subsection{Evaluation of the algorithms performance with 6 players}
In this part, an example is used to illustrate the validity of the differential private distributed NE seeking algorithm that has been designed. Specifically, the topology of the IEEE 30-bus system with 30 buses, 6 generators and 10 local loads (see Fig. 10 in \cite{Wang2004TPS}) is selected to illustrate the effectiveness of the proposed algorithm. Six generators are considered six players, whose time-varying communication topologies are shown in Fig. 1. The time-varying communication graph $\mathcal{T}(l)$ changes periodically from digraphs Fig. 1(1) to Fig. 1(4).

The cost functions are as follows: $J_i(q_i,q_{-i})=q_i(a{\sigma}(q)+b_1)+P_0(\kappa_i(1-\frac{q_i}{b_{2i}})^2+I_i)$, where $\sigma(q)=\frac{1}{6}\sum_{1}^{6}q_i$, $a=0.001$, $b_1=0.1$, $P_0=6$, $b_{21}=2$, $b_{22}=5$, $b_{23}=8$, $b_{24}=12$, $b_{25}=15$, $b_{26}=18$, $\kappa_{1}=5.4$, $\kappa_{2}=4.86$, $\kappa_{3}=4.32$, $\kappa_{4}=4.05$, $\kappa_{5}=3.69$, $\kappa_{6}=4.32$, $I_{1}=0.6$, $I_{2}=0.54$, $I_{3}=0.48$, $I_{4}=0.45$, $I_{5}=0.41$, $I_{6}=0.48$. Then the parameters of the algorithm are designed as follows: $\mu_{i}(l)=\frac{1}{1+0.0001*2^{(0.01l+2)}}$, $\beta_{1}=\beta_{2}=\beta_{3}=\beta_{4}=\beta_{5}=\beta_{6}=0.6$, $\rho(l)=\frac{1}{1+0.1l^{2.01}}$, $b_1(l)=b_2(l)=b_3(l)=b_4(l)=b_5(l)=b_6(l)=2+l$. The initial values are designed as $q(0)=[0.1,0.1,0.1,0.1,0.1,0.1]$, $\hat{w}(0)=[1,1,1,1,1,1]$, $z(0)=[1,1,1,1,1,1]$, $\sigma(0)=[0.1,0.1,0.1,0.1,0.1,0.1]$. And the compact convex sets are $U_1=[-20,20]$, $U_2=[-25,25]$, $U_3=[-30,30]$, $U_4=[-35,35]$, $U_5=[-40,40]$, $U_6=[-45,45]$. The NE can be easily obtained $q_1^{*}=1.9932$, $q_2^{*}=4.9526$, $q_3^{*}=7.8629$, $q_4^{*}=11.6692$, $q_5^{*}=14.4304$, $q_6^{*}=17.2964$.

The simulation results under algorithms (\ref{algorithm01si})-(\ref{algorithm01xi}) are presented in the Fig. 2. The actions of the players under the algorithms (\ref{algorithm01si})-(\ref{algorithm01xi}) of the game problem $\overline{H}$ are shown in Fig. 2(a). The trend of privacy budget at each iteration $l$ is shown in Fig. 2(b). The simulation results show that the actions of the players can arrive at the NE almost surely. For comparison, the differential private distributed NE seeking algorithm designed in \cite{WYQ2024TAC01} under the same noise set and \cite{YEDP2021TAC} under decaying noise parameters are also ran. The convergence error under our algorithm, the algorithm in \cite{YEDP2021TAC}, the Algorithm 1 in \cite{WYQ2024TAC01} are shown with the green, red, and blue curves in Fig. 3, respectively. Thus, our algorithm has a faster convergence speed than that in \cite{WYQ2024TAC01}, and much better accuracy than \cite{YEDP2021TAC}.

\begin{figure}[!t]
\centering
\includegraphics[width=1.8in]{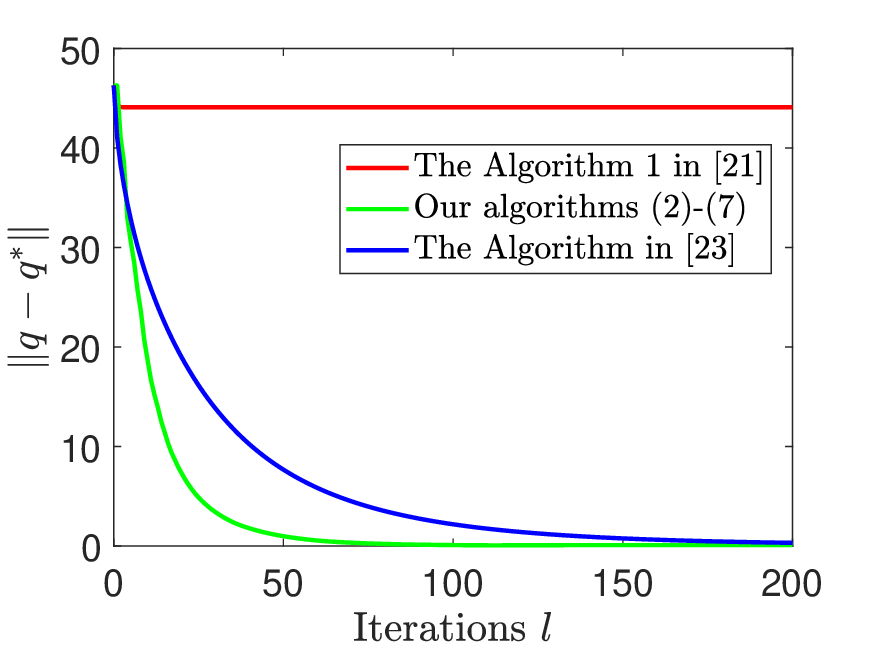}
\caption{Comparison of our algorithms (\ref{algorithm01si})-(\ref{algorithm01xi}), the Algorithm 1 in \cite{YEDP2021TAC}, and \cite{WYQ2024TAC01}.} \label{fig5}
\end{figure}

\subsection{Evaluation of the algorithms performance with diverse network environments in size and density}
In this part, a more comprehensive evaluation of the algorithms performance across diverse network environments is presented.

Firstly, we demonstrate how the algorithm scales with variations in network size under the same noise parameter. The example involving 6, 10, and 17 players are presented to illustrate the performance of the proposed algorithm. The network topologies with 6 players is the same as that in the part A of Simulation, which can be seen in Fig. 1 and changes periodically from digraphs Fig. 1(1) to Fig. 1(4). The IEEE 10-generator 39-bus system (see Fig. 8 in \cite{LI2017TSG}) is selected to describe the time-varying communication topologies with 10 players. The IEEE 162-bus system with 17 generators and 284 transmission lines \cite{OnlineLit} is used to describe the  time-varying communication topologies with 17 players. The convergence error and privacy budget at each iteration of the algorithm with variations in network size are shown in Table I. The corresponding simulation results with convergence error and privacy budget are shown in Fig. 4. From Table I, we can obtain that as the network size increases, the convergence speed of the algorithm becomes increasingly slow. Moreover, the larger the increase in network size, the more pronounced slowdown in convergence speed becomes. With the growth of the network size, the privacy budget at each iteration of the algorithm decreases, which means that the privacy level of the algorithm increases.

Secondly, we demonstrate how the algorithm scales with variations in network density under the same noise parameter. In each iteration, topologies (see Fig. 8 in \cite{LI2017TSG}) with 10 players are considered, with a maximum number of edges of 20, 49, and 79 in all times, and a minimum number of edges of 20, 36, and 50 in all times, respectively. The maximum possible number of edges for 10 players is 90, thus, the density \cite{yang2013TPS} of the three communication graphs are 20/90, 36/90$\sim$49/90 and 50/90$\sim$79/90, respectively. The convergence error and privacy budget at each iteration of the algorithm with variations in network density are shown in Table II. The corresponding simulation results with convergence error and privacy budget are shown in Fig. 5. From Table II, we can obtain that as the network density increases, the convergence speed and the privacy budget at each iteration of the algorithm remain unchanged.

\setlength\arraycolsep{2pt}
\begin{table}[h]
    \centering
    \caption{The convergence error and privacy budget at each iteration of the algorithm with variations in network size under the same noise parameter}
    \begin{threeparttable}
    \begin{tabular}{c|c|c|c}
        \hline
         & 6 \text{Players} & 10 \text{Players} & 17 \text{Players} \\
        \hline
        \text{Iterations}\tnote{1} & 64 & 65 & 89 \\
        \hline
        \text{Privacy budget}\tnote{2} & $3.8655e^{-5}$ & $3.8078e^{-5}$ & $2.8035e^{-5}$ \\
        \hline
    \end{tabular}
    \label{tab:example1}
     \begin{tablenotes}
        \footnotesize
        \item[1] The number of iterations when the convergence error first reaches below 0.5 is considered as ``Iterations''.
        \item[2] Consider the privacy budget value under the ``Iterations''.
      \end{tablenotes}
  \end{threeparttable}
\end{table}

\setlength\arraycolsep{2pt}
\begin{table}[h]
    \centering
    \caption{The convergence error and privacy budget at each iteration of the algorithm with variations in network density  under the same noise parameter (10 players)}
    \begin{threeparttable}
    \begin{tabular}{c|c|c|c}
        \hline
         & 0.222 \text{Density}\tnote{3} & 0.544 \text{Density} & 0.878 \text{Density}\\
        \hline
        \text{Iterations}\tnote{1} & 65 & 65 & 65 \\
        \hline
        \text{Privacy budget}\tnote{2} & $3.8078e^{-5}$ & $3.8078e^{-5}$ & $3.8078e^{-5}$ \\
        \hline
    \end{tabular}
    \label{tab:example2}
    \begin{tablenotes}
        \footnotesize
        \item[1] Same as Table I.
        \item[2] Same as Table I.
        \item[3] ``0.222 Density'' represents the maximum density value of the graph at all times is 0.222. Others are similar.
      \end{tablenotes}
  \end{threeparttable}
\end{table}

% The convergence and privacy budget at each iteration of the algorithm with variations in network size
\begin{figure}[t]
\centering  %图片全局居中
\subfigure[Convergence error.]{
\label{Fig.sub.1}
\includegraphics[width=1.6in]{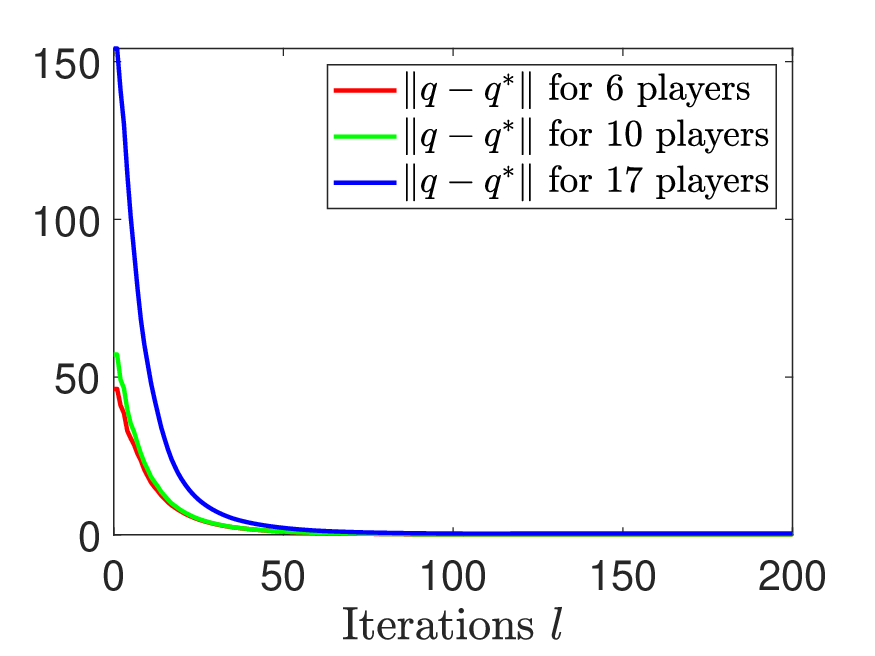}}
\subfigure[Privacy budget.]{
\label{Fig.sub.2}
\includegraphics[width=1.6in]{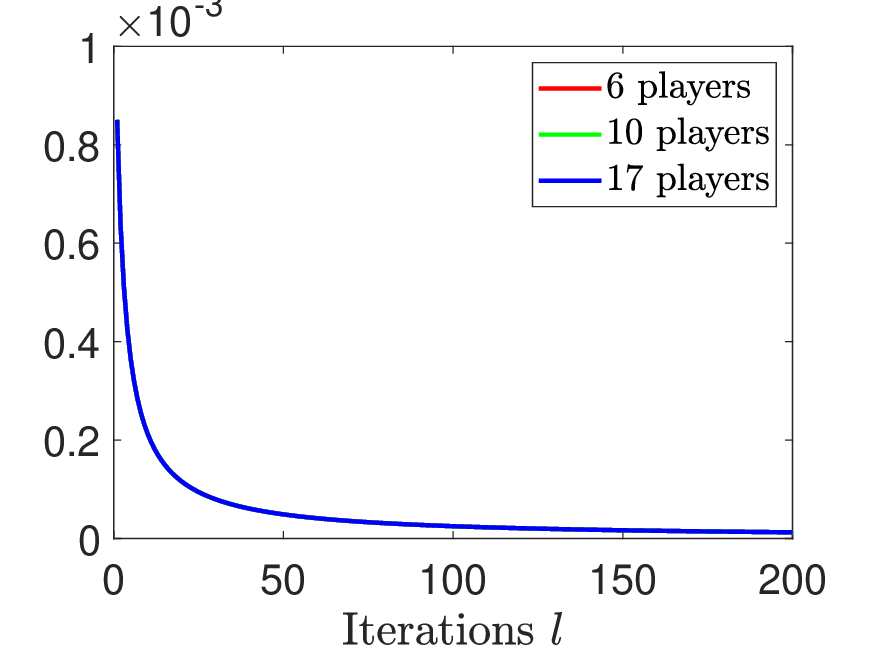}}
\caption{The convergence error and privacy budget at each iteration of the algorithm with variations in network size.}
\label{size01}
\end{figure}

% The convergence rate and privacy budget at each iteration of the algorithm with variations in network density (10 players)
\begin{figure}[t]
\centering  %图片全局居中
\subfigure[Convergence error.]{
\label{Fig.sub.1}
\includegraphics[width=1.6in]{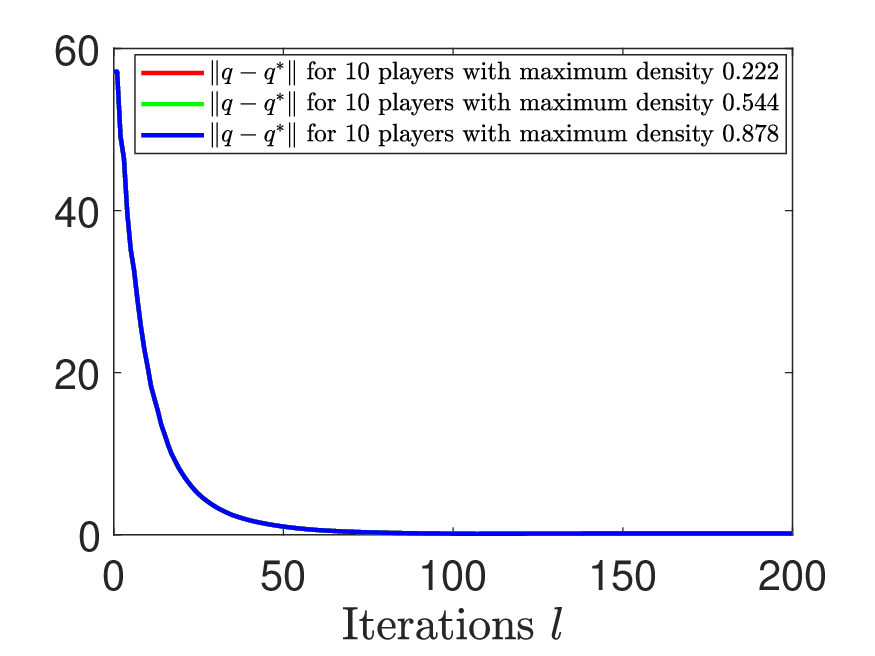}}
\subfigure[Privacy budget.]{
\label{Fig.sub.2}
\includegraphics[width=1.6in]{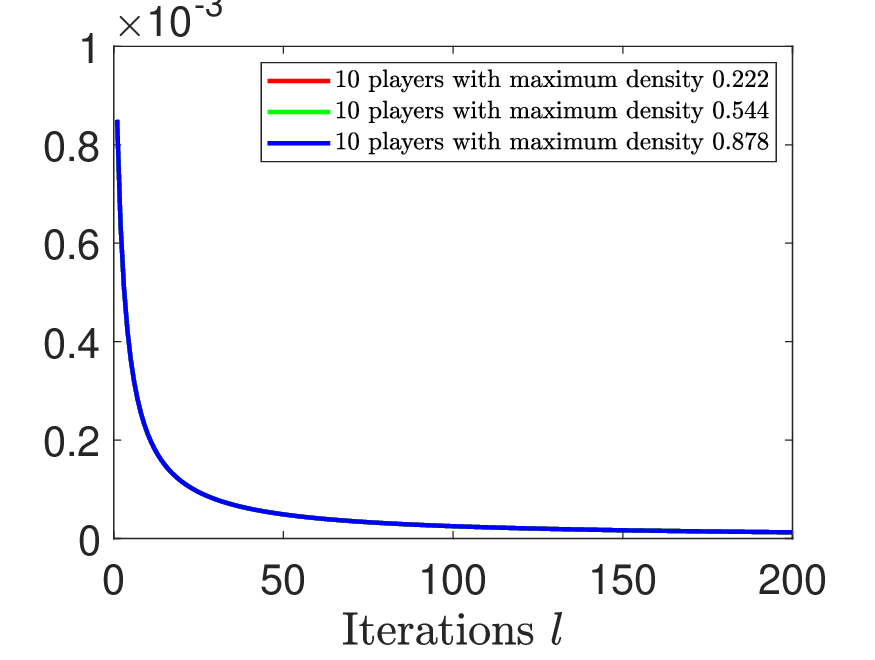}}
\caption{The convergence error and privacy budget at each iteration of the algorithm with variations in network density.}
\label{density01}
\end{figure}

\section{Conclusion}
The differential private distributed NE seeking problem in aggregative games on time-varying unbalanced directed communication graphs is explored. For protecting the privacy of players' sensitive information, the random independent noises drawn from Laplace distributions are injected into the transmitted information in order to mask the real information. Then the push-sum consensus protocol is designed to alleviate network imbalance for estimating the aggregate function by using perturbed information. To attenuate the negative affect of the noise, the weakening factor is utilized to design differential private distributed NE seeking algorithm combining the momentum term, the gradient descent and projection-based method to ensure the convergence of the algorithm. The simulation results are presented to show the validity of the algorithm, which shows that the actions of the players can almost surely converge to NE, while achieving rigorous differential privacy with a summable cumulative privacy budget without requiring the trade-off between provable convergence and differential privacy.

\balance
\bibliographystyle{plain}        % Include this if you use bibtex

\end{document}